\documentclass[prb,twocolumn,preprintnumbers,amsmath,amssymb]{revtex4}
\usepackage{graphicx}
\usepackage{epsfig}
\usepackage{amsmath,amscd}
\usepackage{psfrag}
\usepackage[dvipdfm]{hyperref}
\usepackage{amsmath,amscd}
\usepackage{psfrag}

\begin{document}

\title{Effect of vortex hotspots on the radio-frequency surface resistance of superconductors.}
\author{A. Gurevich$^1$}\email{gurevich@odu.edu}
\author{G. Ciovati$^2$}\email{gciovati@jlab.org}
\affiliation{$^1$Department of Physics and the Center for Accelerator Science, Old Dominion University,
Norfolk, VA 23529}
\affiliation{$^2$Thomas Jefferson National Accelerator Facility, Newport News, VA 23606}

\begin{abstract}
We present detailed experimental and theoretical investigations of hotspots produced by trapped vortex bundles and their effect on the radio-frequency (rf) surface resistance $R_s$ of superconductors at low temperatures. Our measurements of $R_s$ combined with the temperature mapping and laser scanning of a 2.36 mm thick Nb plate incorporated into a 3.3 GHz Nb resonator cavity cooled by the superfluid He at 2 K, revealed spatial scales and temperature distributions of hotspots and showed that they can be moved or split by thermal gradients produced by the scanning laser beam. These results, along with the observed hysteretic field dependence of $R_s$ which can be tuned by the scanning laser beam, show that the hotspots in our Nb sample are due to trapped vortex bundles which contain $\sim 10^6$ vortices spread over regions $\sim 0.1-1$ cm. We calculated the frequency dependence of the rf power dissipated by oscillating vortex segments trapped between nanoscale pinning centers, taking into account all bending modes and the nonlocal line tension of the vortex driven by rf Meissner currents. We also calculated the temperature distributions caused by trapped vortex hotspots, and suggested a method of reconstructing the spatial distribution of vortex dissipation sources from the observed temperature maps. Vortex hotspots can dominate the residual surface resistance at low temperatures and give rise to a significant dependence of $R_s$ on the rf field amplitude $H_p$, which can have important implications for the rf resonating cavities used in particle accelerators and for thin film structures used in quantum computing and photon detectors.

\end{abstract}


\maketitle

\section{introduction}
The physics of electromagnetic response of superconductors at low temperatures has recently attracted much attention due to its importance
for the understanding of the behavior of resonator cavities for particle accelerators \cite{cav,cav1,agac} micro resonating striplines \cite{mres} and other superconducting thin film structures used for qubits \cite{qcomp,qq} and photon detectors \cite{det1,det2,det3}. One of the main parameters of merit of such structures is the quality factor $Q=G/R_s$ inversely proportional to the surface resistance $R_s$, where $G\simeq g_0\mu_0c$, $\mu_0c=377$ $\Omega$ is the vacuum impedance, $c$ is the speed of light, and $g_0\sim 1$ is a geometric factor. For s-wave superconductors, the BCS and Eliashberg theories \cite{mb,agh,nam} predict an exponentially small
$R_s(T)$ in the Meissner state at low rf frequencies $\omega \ll k_BT_c/\hbar$ and temperatures $T\ll T_c$ much lower than the critical temperature $T_c$. However, $R_s(T)$ observed on many s-wave superconducting materials is usually described by \cite{turn,hein,gigiJAP}
\begin{equation}
R_s=(A\omega^2/T)\exp(-\Delta/k_BT)+R_i,
\label{rbi}
\end{equation}
where the first term is the BCS contribution due to thermally-activated quasiparticles, and $\Delta$ is the superconducting gap. In the dirty limit the Mattis-Bardeen theory yields $A\simeq (\mu_0^2\lambda^3\Delta/\rho_nk_B)\ln(9k_BT/4\hbar\omega)$, where $\lambda$ is the London penetration depth, and $\rho_n$ is the normal state resistivity\cite{mb,mres}. The last term in Eq. (\ref{rbi}) is the residual resistance $R_i$ which remains finite as $T\rightarrow 0$. The Nb resonator cavities can reach $Q\simeq 10^{10}-10^{11}$ and very low $R_s\simeq 10-30$ n$\Omega$ and $R_i\simeq 2-10$ n$\Omega$ at 2 K \cite{cav,agac}.

Residual resistance can be a significant source of the rf dissipation in resonator cavities and superconducting thin film qubits at very low temperatures. Generally, nonzero $R_i$ implies a finite density of subgap states at the quasiparticle energies $|E|<\Delta$ and a finite density of states at the Fermi level, as was indeed revealed by tunneling measurements \cite{dynes,prosl,hoffman}. The subgap states have been attributed to inelastic scattering of electron on phonons \cite{inelast}, strong Coulomb correlations \cite{coulomb}, local variations of the BCS coupling constant by impurities \cite{larkin,feigel} or pairbreaking magnetic impurities taken into account in a more rigorous way than in the original Abrikosov-Gor'kov theory \cite{balatski}. In other theories the tail in $N(E)$ at $E < \Delta$ results from spatial correlations in impurity scattering \cite{larkin,meyer}.

Besides the subgap states, $R_i$ has also been attributed to such extrinsic factors as grain boundaries \cite{gb1,gb2,gb3,gb4}, nonsuperconducting second phase precipitates \cite{cav1}, or generation of hypersound by the rf field \cite{sound}. Another significant contribution to $R_i$ comes from trapped vortices oscillating under the rf field \cite{gr,rabin,cc,gurci,vortrf}. Trapped vortices can appear during the cooldown of a superconductor through $T_c$ due to the effect of stray magnetic fields $H>H_{c1}(T)$, including the unscreened Earth field, since the lower critical field $H_{c1}(T)$ vanishes at $T_c$. This mechanism becomes particularly important in thin films where vortices can be generated by very weak perpendicular stray fields as the perpendicular $H_{c1}$ is reduced by a large demagnetizing factor \cite{ehb}. Spontaneous vortex-antivortex pairs and vortex loops can appear upon cooling a superconductor with a finite temperature ramp rate \cite{kz1,kz2,kz3,kz4,kz5,kz6,kz7}, or be produced by thermal fluctuations \cite{kz3} even if a superconductor is fully screened against external magnetic fields. Generation of trapped vortices due to these very weak mechanisms is negligible in typical magnetization measurements of superconductors, but it can give the main contribution to the exponentially small surface resistance in the Meissner state at very low temperatures. This  contribution becomes apparent in the resonator rf cavities because of their extremely high quality factors $Q\sim 10^{10}-10^{11}$ at 2 K, so the Nb cavities in which the densities of screening Meissner currents can reach the depairing limit \cite{cav1,agac}, can be a unique tool to probe the dynamics of mesoscopic vortex structures under strong rf fields.

Trapped vortices in random pinning potential can bundle together, forming localized hotspots in which vortices oscillate under the rf field. Hotspots in the Nb resonator cavities have been revealed by temperature map measurements using arrays of carbon thermometers mounted at the outer cavity surface \cite{gigiJAP,tmap}. Hotspots due to trapped vortices have also been observed on thin film structures \cite{vortrf,anlage}. Given that many materials factors such as inhomogeneous distribution of impurities, lossy nonsuperconducting precipitates, grain boundaries, surface topography and other structural defects can also result in hotspots in resonator cavities \cite{cav,cav1,agac}, or THz radiation sources based on the layered cuprates \cite{thz1,thz2}, distinguishing vortex hotspots from hotspots caused by materials defects becomes important. This can be done using the fact that vortices, unlike the hotspots due to fixed materials defects, can be moved by thermal gradients \cite{agg} produced by outside heaters \cite{gigih} or scanning laser beams \cite{gigil,asc12}. Thus, any changes in the temperature maps observed before and after applying thermal gradients would indicate that the underlying hotspots are due to trapped vortices.

The unprecedented sensitivity of the Nb resonator cavities with $Q\sim 10^{10}-10^{11}$ combined with the temperature mapping and the scanning laser techniques give an opportunity to probe the behavior of low density vortex structures which can give rise to the observed residual resistance under the rf field. This situation can also be relevant to other high-$Q$ structures such as thin film superconducting qubits \cite{qcomp,qq}, photon detectors \cite{det1,det2,det3} or superconducting screens used in the search of magnetic monopoles \cite{monopole}. Moreover, the laser \cite{lsm1,lsm2} and electron beam \cite{esm1,esm2} scanning techniques can be used not only to move vortex hotspots, but also annihilate and break them into pieces by increasing the beam intensity. Displacement of vortices by scanning electron beams has been demonstrated in annular Josephson junctions \cite{ustin} and thin film SQUIDs \cite{esquid1,esquid2}, and also calculated theoretically \cite{clem}.

The behavior of vortex hotspots under the rf field is related to the following outstanding issues: 1. Power dissipated by trapped vortices and its dependence on the rf frequency and the geometry of the pinned vortex segments. 2. Temperature distributions produced by oscillating vortex segments and their detection by temperature map experiments, 3. Moving and breaking vortex hotspots by scanning laser beams and possibilities of using this technique to reduce the contribution of vortices to rf dissipation, 4. Contributions of trapped vortex hotspots to the low-field residual resistance, as well as the nonlinear surface resistance at high rf fields at which the hotspots start expanding and can ignite thermal quench propagation. In this paper we address these issues by combining the surface resistance and temperature map measurements, scanning laser technique and theory. Our experiments were performed on a Nb plate incorporated in a resonating cavity. The paper is organized as follows.

In Section II we discuss mechanisms by which various vortex configurations can be trapped upon cooling a superconductor and the ways by which these
vortices can be moved by thermal gradients produced by scanning laser beams. Section III describes the temperature mapping technique which was used to reveal
hotspots in the Nb plate mounted inside a Nb resonator cavity. We show that scanning the surface of the Nb plate with a laser beam moves and breaks hotspots and causes hysteretic behavior of the surface resistance on the rf field, indicating that these hotspots are indeed caused by trapped vortices. In Section IV we present detailed calculations of the rf power produced by single vortices as functions of the rf frequency and the length of a pinned vortex segment. Section V describes calculations of temperature distributions produced by vortex hotspots and reconstruction of the underlying dissipation sources from the measured temperature maps. In Section VI we address the effect of trapped vortices on the residual resistance and the nonlinear surface resistance at high rf fields. Section VII contains discussion of the results.

\section{Trapped vortices in superconductors}

\subsection{Generation of trapped vortices}

Trapped vortices can be produced by any external magnetic field $H>H_{c1}(T)$ upon cooling a superconductor through $T_c$. For instance, the unscreened Earth field $B_E=\mu_0H_E\simeq 40\ \mu$T can generate vortices spaced by $a\simeq (\phi_0/B_E)^{1/2}\simeq 7\ \mu$m. Since $H_{c1}(T)\simeq H_{c1}(0)(1-T^2/T_c^2)$ increases as $T$ decreases ($B_{c1}(0)\simeq 170$ mT for Nb), the subsequent cooldown to lower temperatures at which $H\ll H_{c1}(T)$ makes vortices thermodynamically unstable, forcing them to escape through the sample surface. In doing so a fraction of vortices can get trapped by the materials defects such as non-superconducting precipitates, networks of dislocations or grain boundaries, giving rise to pinned vortex bundles depicted in Fig. 1. In the field cooled state vortices are mostly oriented along the shortest sample dimension, since the perpendicular $H_{c1}^\bot \sim H_{c1}d/w$ is strongly reduced by the demagnetizing factor of films with small aspect ratio $d/w\ll 1$, where $d$ and $w$ is the film thickness and width, respectively \cite{ehb,geshk}. Despite a seemingly weak effect of the Earth field, it can result in the rf vortex dissipation exceeding the BCS contribution in the Meissner state at $T\ll T_c$, so the Nb accelerator cavities are magnetically screened to reduce the Earth magnetic field by $\sim 10-100$ times \cite{cav}.

Even the complete magnetic screening cannot fully suppress the formation of trapped vortices, particularly vortex loops which appear
spontaneously during the cooldown through $T_c$ and then get trapped by pinning materials defects. This can occur, for example, due to the
Kibble-Zurek mechanism \cite{kz1,kz2} of generation of vortex-antivortex pairs with the areal density $n_f\sim (\tau_{GL}/\tau_Q)^{1/2}\xi_0^{-2}$ in thin films. Here $\tau_{GL}\simeq \pi\hbar/8k_BT_c$ is the characteristic relaxation time of the
superconducting order parameter, the time $\tau_Q = T_c/(dT/dt)$ quantifies the temperature cooling rate $dT/dt$, and $\xi_0$ is the coherence length
at $T=0$. The vortex density generated by this mechanism would be equivalent to the magnetic field,
\begin{equation}
B_{KZ}\sim B_{c2}(\tau_{GL}/\tau_Q)^{1/2},
\label{KZ}
\end{equation}
where $B_{c2}=\phi_0/2\pi\xi_0^2$ is the upper critical field, and $\phi_0$ is the magnetic flux quantum. For Nb with $T_c=9.2$ K, $B_{c2}\simeq 0.4$ T, $\tau_{GL}\sim 10 ^{-12}$ s, and $\tau_Q\sim 1$ s, Eq. (\ref{KZ}) predicts $B_{KZ}\simeq 0.4\ \mu$T. As will be shown below, even such small fields
$B_{KZ}\ll B_E$ could result in a residual surface resistance $R_i\sim 1$ n$\Omega$, of the order of what has been observed on the high-performance Nb cavities.

The Kibble-Zurek scenario has been tested experimentally on superconducting films cooled down with different rates \cite{kz3,kz4,kz5}, and by numerical simulation of the time-dependent Ginzburg-Landau equations \cite{kz6,kz7}. While the experiments have shown spontaneous generation of vortex-antivortex pairs in superconducting films cooled down through $T_c$ with different rates, significant numerical discrepancies with the model \cite{kz1,kz2} have been observed.
Alternatively, it was proposed that vortex loops can be generated by thermal fluctuations in a narrow temperature region near $T_c$ where the line energy of the vortex $\epsilon \propto 1-T/T_c$ vanishes \cite{kz3}. Suggestions that trapped vortices can be generated by magnetic fields caused by thermoelectric currents \cite{cav} seem less plausible since thermoelectric currents vanish in the Meissner state \cite{thermoel}.

In this work we address the effect of trapped vortices on the surface resistance irrespective of particular mechanisms by which vortices can appear in a zero-field cooled superconductor. Usually these weak mechanisms produce low-density vortex structures in which the vortex spacing is larger than $\lambda$ so that interaction between vortices can be neglected. We assume that vortices are trapped by randomly-distributed materials defects, giving rise to distorted vortex lines which can either connect the opposite faces of the sample or form closed loops in the bulk or semi-loops ending on one side of the sample, as depicted in Fig. 1. For thick film screens with $d\gg \lambda$, only short tips of these pinned vortex segments are exposed to the rf field.
%\bigskip

\begin{figure}
\centerline{\psfig{file=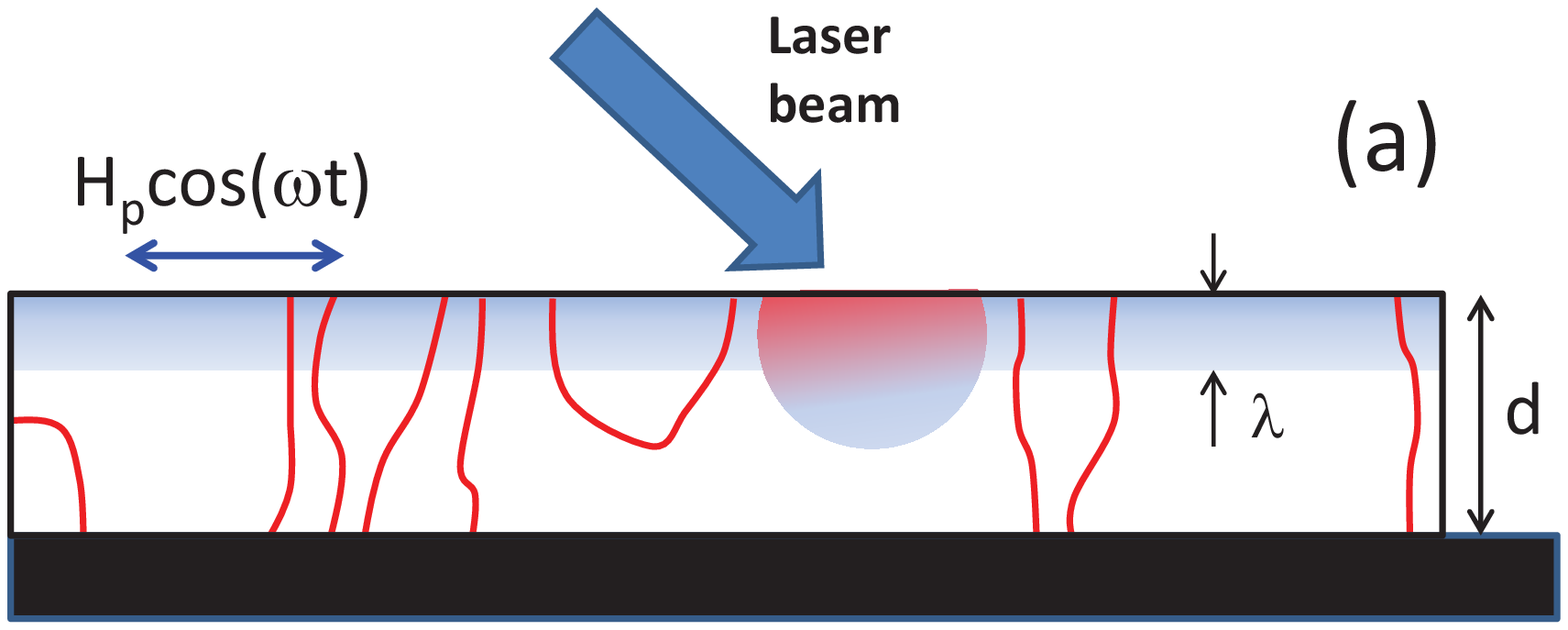,width=6.5cm}}
\bigskip
\centerline{\psfig{file=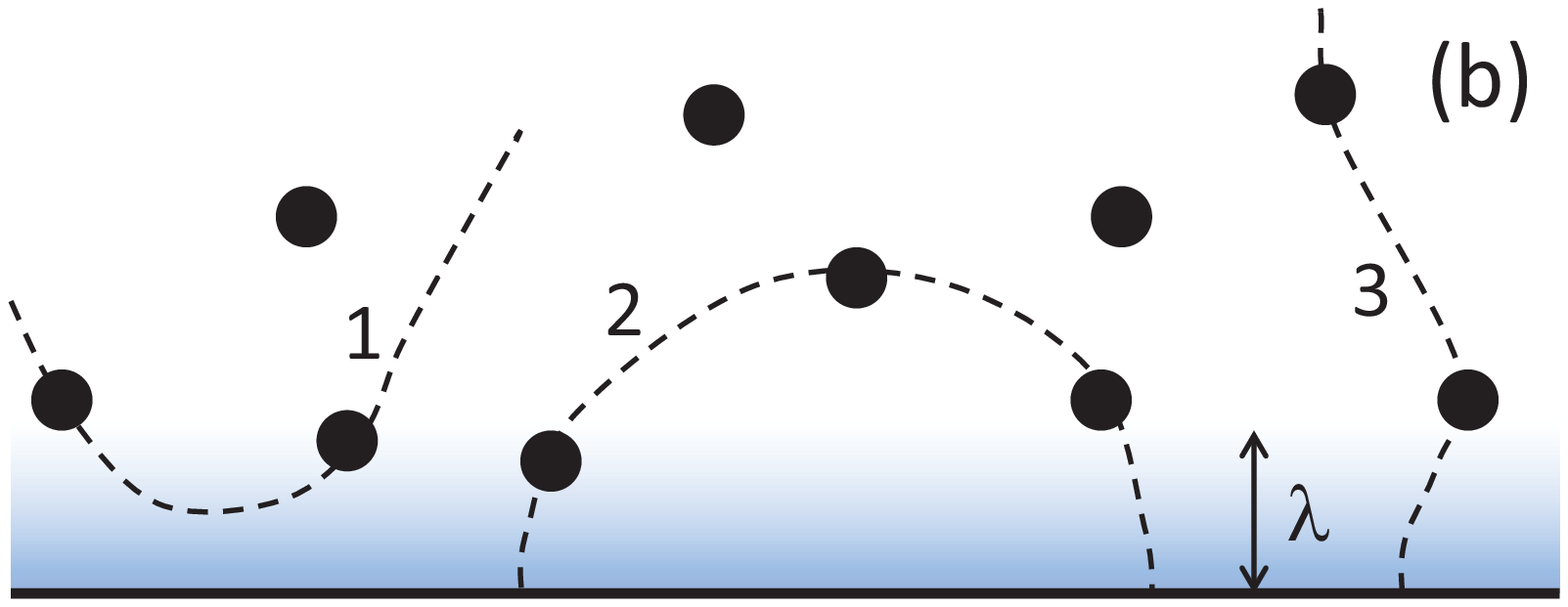,width=5.5cm}}
\caption{Trapped vortices in a film on a substrate (black), the top of the film being exposed to the rf field penetrating in the
 layer of thickness $\sim \lambda$. A scanning laser beam produces a moving hotspot which can push the tips of the vortex lines along the film (a). Possible vortex configurations at the surface exposed to the rf field (b): A vortex segment crossing the region of the rf field penetration (1), Pinned vortex semi-loop at the surface (2), Vortex line going all the way across the film (3). The black regions depict pinning centers.}
\label{fig1}
\end{figure}

Figure 1 shows three different types of vortex configurations contributing to $R_i$: 1. small segments of a vortex line close to the surface, 2. vortex loops starting and ending at the surface exposed to the rf field, 3. vortex lines connecting two opposite faces of the sample. Here only tips of vortices perpendicular to the surface or fraction of parallel vortex segments spaced by $\lesssim 2\lambda$ from the surface are exposed to the rf field. Calculation of $R_i$ for a pinned flexible vortex segment parallel to the surface (case 1 in Fig. 1b) was done before \cite{gurci}. Here we mostly focus on cases 2 and 3 for which the rf dissipation in a thick film $(d\gg \lambda)$ is mostly determined by the distribution of oscillating vortex tips along the surface and the lengths of the dangling vortex segments $\ell$ defined by the distance between the nearest pinning center and the surface, as illustrated by Fig. 1b. We are considering here mesoscopic vortex bundles oscillating under the rf field and producing hotspots which are then detected by an array of thermometers on the sample surface.

\subsection{Moving vortices by thermal gradients.}

Vortices can be moved by the Lorentz force produced by the superfluid current density $\bf{J}$ or by the thermal force ${\bf f}_T=-s^*\nabla T$ exerted by the temperature gradient $\nabla T$, (Ref. \onlinecite{huebener}). Here $s^*(T)\simeq-\beta_c\partial\epsilon_0/\partial T$ is the transport entropy carried by quasi-particles in the vortex core, $\beta_c\epsilon_0$ is the free energy of the vortex core, $\epsilon_0=\phi_0^2/4\pi\mu_0\lambda^2$, and $\beta_c\approx 0.38$ is evaluated numerically from the GL equations \cite{core}. Vortex segments can be shifted from one pinned configuration to another if $f_T$ locally exceeds the pinning force $\phi_0J_c$ per unit length where $J_c$ is the critical current density. The depinning temperature gradient $|\nabla T|_c$ is estimated from $s^*|\nabla T|_c = \phi_0 J_c$:
\begin{equation}
|\nabla T|_c\simeq \phi_0J_c/s^*\sim 16\mu_0J_c\lambda_0^2T_c^2/\phi_0 T,
\label{nT}
\end{equation}
where we used $\lambda^{-2}(T)=\lambda_0^{-2}(1-T^2/T_c^2)$. Taking here $\lambda_0 = 40$ nm, $T_c=9.2$ K, and $J_c\sim 10^7$ A/m$^2$ for clean Nb, (Ref. \onlinecite{JcNb}), yields $|\nabla T|_c\sim 7$ K/mm at 2 K.

Vortex hotspots can be moved or split by thermal gradients caused by outside heaters, as was observed on the Nb resonator cavities \cite{gigih}. In this work we use a scanning laser beam to produce a moving hot region (in which the temperature can locally exceed $T_c$) to depin trapped vortex segments.  This method is basically a higher-power version of the scanning laser \cite{lsm1,lsm2} or electron beam \cite{esm1,esm2} microscopy, which allows us not only to probe vortex hotspots but also to selectively apply high temperature gradients to a particular vortex bundle and push it in a desired direction. The effect of temperature gradient depends on particular configurations of pinned vortices shown in Fig. 1. For instance, a hot region produced by the laser beam can depin the vortex segment 1 and push it away from the surface by a distance $\gg \lambda$, where this segment gets trapped by another pin and can no longer be reached by the rf Meissner currents. Likewise, the laser beam can depin and annihilate the loop 2 in Fig. 1b but it can only shift a tip of vortex 3 along the surface. Therefore, vortices connecting the opposite faces of a flat sample cannot be eliminated by thermal gradients, unless they are pushed all the way to the sample edges. Yet the laser beam can depin the ends of threading vortices which are then  trapped by other pins. As a result, the vortex ends get redistributed along the surface and can either spread around or clump together, depending on a particular configuration of pins. Experimental evidence of these effects will be shown below.

\section{ EXPERIMENTAL SETUP AND MEASUREMENTS}

We have developed an experimental setup which allows us to measure the surface resistance and to scan a laser beam with adjustable size and power onto the horizontal inner surface of a semi-spherical Nb cavity placed in a vertical cryostat (2.75 m high and 71 cm in diameter) filled with a superfluid He, as shown in Fig. 2. Technical details of this setup are described elsewhere \cite{gigil}. The same apparatus was used to obtain maps of the surface resistance using a low-temperature laser scanning microscopy (LSM) technique \cite{gigil}. A 10 W, 532 nm continuous-wave laser is placed on the top plate of the cryostat, along with optical components which allow adjusting the output power between $\sim 3$ mW and 9.8 W and the beam diameter, defined at $1/e^2$ of the maximum intensity, at the cavity location between 0.87 mm and 3.0 mm. Two remotely-rotatable scanning mirrors are located inside a vacuum chamber on top of the cavity and allow scanning of the laser beam in the x-y direction.

%\bigskip
\begin{figure}
\centerline{\psfig{file=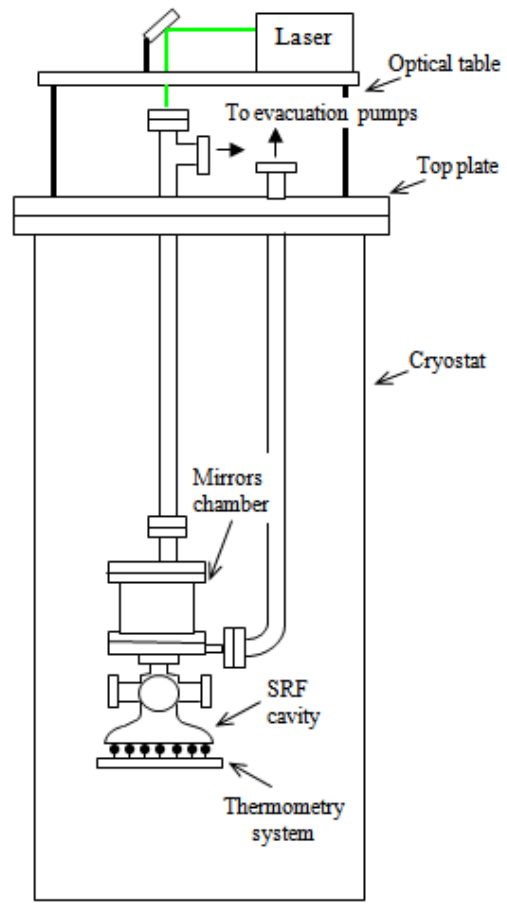,width=6.5cm}}
\caption{Schematic of the experimental setup used for local heating of the inner surface of the Nb cavity with a scanning laser beam \cite{gigil}.}
%\label{fig2}
\end{figure}

The Nb cavity consists of a half-cell of the TESLA shape \cite{cav} with a flat Nb plate of diameter $20.7$ cm and thickness $d=2.36$ mm welded at the equator. The resonant frequency of the TM$_{010}$ mode was 1.3 GHz, however in our experiments we excited a different resonant mode: the TE$_{011}$ mode at 3.3 GHz because it ideally has zero surface electric field, thereby minimizing the field emission of secondary electrons \cite{cav}. In the TE$_{011}$ mode the magnitude of the rf magnetic field $B_\|(r)$ parallel to the surface vanishes in the center and at the edges of the flat Nb plate, $B_\|(r)$ increasing from 60 to 90$\%$ of the peak value $B_p$ as the radial distance $r$ from the center increases from $10$ mm to $35$ mm, respectively. Our setup allows deflecting the laser beam to the maximum distance $r_m\approx 40$ mm from the center of the Nb plate. The peak surface magnetic field occurs in a region near the cavity iris, which is not accessible by the laser. The geometry factor, $G=Q_0R_s$, of the TE$_{011}$ mode in the cavity is 501.2 $\Omega$. The cavity was built from a large-grain (a few mm grain size) Nb with the residual resistivity ratio of $\rho(300K)/\rho(T_c)\simeq 200$. To avoid contamination of the surface of the Nb plate during measurements, the laser beam was transmitted through an optical window which isolated the cavity from the vacuum chamber with the scanning mirrors.

The post-fabrication cavity treatment consisted of $~100\ \mu$m material removal from the inner surface by buffered chemical polishing (BCP) with HF:HNO$_3$:H$_3$PO$_4$ = 1:1:2, heat treatment in an ultra-high vacuum (UHV) furnace at $800^\circ$C for 3 h followed by additional BCP to remove $\sim 20\ \mu$m damaged layer from the inner surface. The typical surface preparation includes the conventional high-pressure water rinse with ultra-pure water \cite{cav}, assembly of input and pick-up rf antenna, attachment to the mirrors chamber on a vertical test stand and evacuation.

Another key component of the setup is an array of thermometers attached to the outer surface of the flat Nb plate. The system consists of 128 calibrated resistance temperature sensors evenly distributed along seven concentric "rings" of radii from 12.5 to 88.9 mm, such that the distance between neighboring sensors is 12-20 mm. The thermometry system allows identifying the rf hotspots as well as locating the position of the laser beam on the cavity flat plate and verifying the movement of the beam during laser sweeping.

The experiment proceeded as follows: (a) the cavity was cooled-down to 2 K, (b) a baseline rf measurements were performed and the thermometry was used to identify hotspots on the flat Nb plate, (c) the rf power was switched off, the laser beam was turned on and directed to each hotspot and then swept with different patterns. (d) The laser was turned off, the rf power was switched on and another rf measurement was performed using thermometry to detect changes in the temperature maps. Different sweeping patterns, such as line scans along the $x$ or $y$ axis, inward and outward spiral laser scanning have been tried, given that neither distribution of pinning centers nor orientation of trapped vortices are known in advance. For example, if a vortex is pinned at a grain boundary (GB), sweeping the laser in different directions would reveal the grain boundary orientation, depending on the direction the vortex would move. If intragrain pinning is due to random defects, GBs become
channels of preferential motion of vortices for which pinning along GB is weaker than pinning in the perpendicular direction \cite{gb}.  Furthermore, using an inward spiral trajectory of the laser beam, starting at a large radius from the cavity center and ending at the cavity center, it might be possible to drag vortices towards the center of the cavity, where the amplitude of the surface magnetic field is close to zero. In the latter case the scanning laser is used as a "thermal broom"  which can reduce the global rf dissipation in the cavity.

\section{EXPERIMENTAL RESULTS}

We show here representative results of many measurements based on the procedure outlined in the previous section (some earlier data were published in \cite{asc12}). The first group of measurements (labeled "test No. 11c") were obtained after the cavity had $\simeq 27\ \mu$m additional material removal by BCP 1:1:2, heat treated in UHV at $600^\circ$C for 10 h, then etched for additional $\simeq 9\ \mu$m material removal. The cavity wall thickness at several locations was measured after each chemical etching step with an ultrasonic probe. For test No. 11c, a solenoid (20 mm in diameter and 50 mm long) was co-axially mounted at $\simeq 25$ mm below the Nb plate, underneath the thermometry system. During the first cool-down to 2 K the residual resistance $R_i\simeq 30 \pm 8$ n$\Omega$ was obtained from the Arrhenius plot of $R_s(T)$ between 4 K and 2 K. Then the cavity was warmed up to 20 K, the solenoid was powered up to generate a maximum dc field $\simeq 13.8\ \mu$T at the Nb plate as the cavity was cooled down again. Once the cavity temperature reached 4.3 K, the solenoid was turned off and the temperature of the cavity was lowered to 2.0 K by pumping on the liquid He bath inside the cryostat. The residual resistance increased to $\simeq 150$ n$\Omega$, resulting in $Q_0\simeq 2.3\cdot 10^9$. The rf power was increased for the baseline test and a brief processing of multipacting (MP) - the field emission of electrons which then produces an avalanche of secondary electrons repeatedly impacting the Nb surface, occurred above $B_p \simeq 40$ mT. This and another MP onset at $B_p \simeq 70$ mT were suppressed by He processing. Quenches were observed at $B_p \simeq 92$ mT but the quench location was not on the Nb plate.

%\bigskip
\begin{figure} [ht]
\includegraphics[width=\columnwidth]{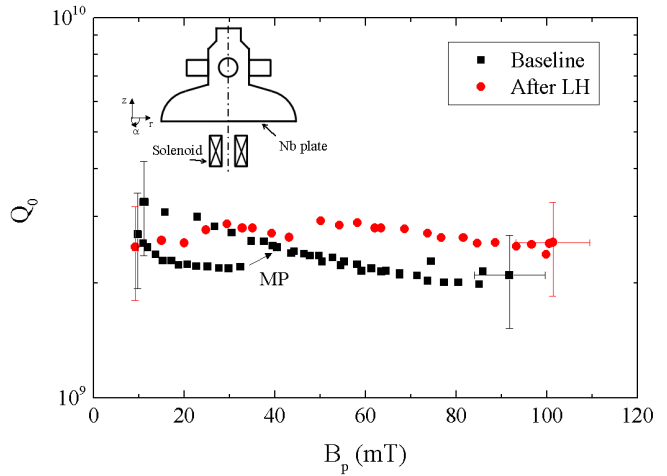}
\caption{$Q_0(B_p)$ measured at 2.0 K during test No. 11c, before and after laser heating. Insert shows the
geometry of the dc magnet relative to the Nb plate.}
\label{fig3}
\end{figure}
\bigskip
\begin{figure}
\includegraphics[width=\columnwidth]{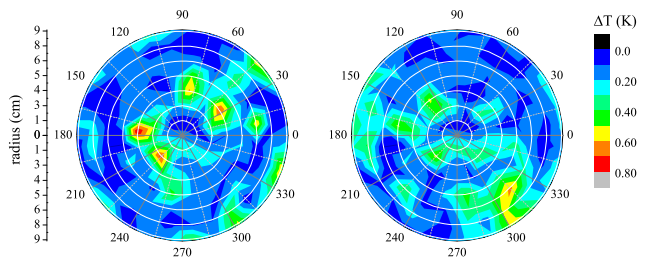}
\caption{Temperature maps at 2 K, $B_p = 74$ mT measured in test 11c while decreasing the rf field
during the baseline test (left) and while increasing the rf field after laser heating (right).}
\label{fig4}
\end{figure}

The $Q_0(B_p)$ curve for the baseline test is shown in Fig. 3. The rf field was then reduced back to 10 mT and the rf power was switched off. Shown in Fig. 4 is the temperature map of the Nb plate measured at $B_p = 74$ mT as the rf field was ramped down. Hotspots are clearly present on rings 2 and 3 (rings of thermometers are numbered from 1 to 7 for increasing ring diameter). The laser beam with the diameter of 0.87 mm and power of 10 W was directed to four locations of the Nb plate on rings 2 and 3 (indicated by arrows in Fig. 5) and a sweep following an outward spiral trajectory was done starting from the center of each hotspot location. The maximum radius of the spiral trajectory was 1 cm, changed with 1 mm increment. The laser was then turned off and a new rf measurement of $Q_0(B_p)$ showed that the cavity quenched at $B_p \simeq 100$ mT with higher $Q_0$, as evident from Fig. 3. Furthermore, Fig. 4 shows that the temperature map of the hotspot distribution measured at $B_p = 74$ mT changed after laser heating (LH). Such changes are also evident from the profiles of the local temperature rise, $\Delta T$, measured by thermometers along rings 2 and 3 before and after the laser sweep presented in Fig. 5. These results clearly show that, as a result of laser scanning, hotspots do move and reduce or increase their intensity. Here the true values of $\Delta T({\bf r})$ at the outer surface of the Nb plate can be estimated by dividing the measured $\Delta T$ by the thermometers' efficiency, which is about $\sim 20\%$. Some of the sensors turned out to be located at grain boundaries (marked as "GB" in Fig. 5) observed by optical microscopy of the Nb plate. Temperature maps measured before and after LH indicate that grain boundaries do not necessarily manifest themselves as rf hotspots relative to other places of the Nb sample.

\begin{figure}
\includegraphics[width=7.5cm]{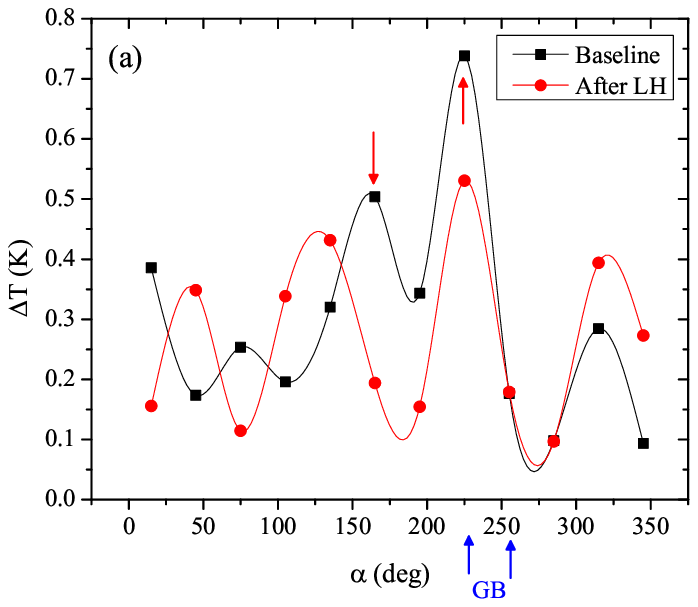}
\includegraphics[width=7.5cm]{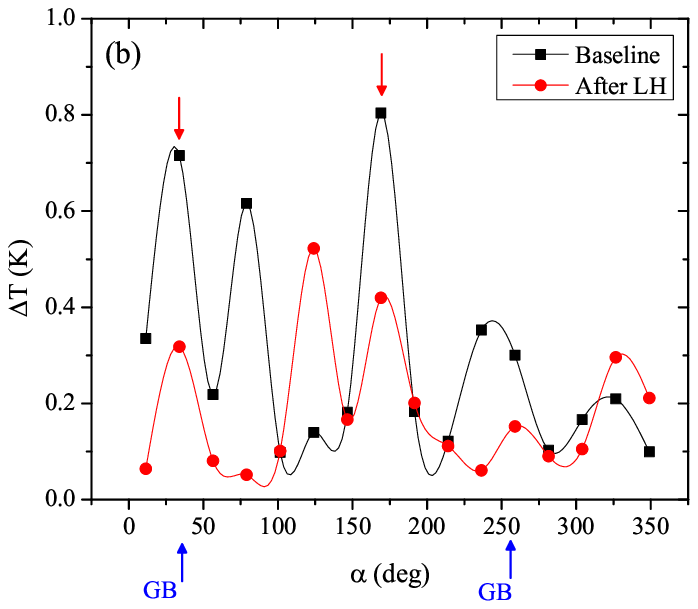}
%\bigskip
\caption{Temperature rise $\Delta T$ measured at 2.0 K, 74 mT by thermometers along rings 2 (a) and 3 (b) in the temperature maps of Fig. 4, before and after laser heating in test No. 11c. The red arrows indicate the locations where the laser beam was scanned following an "outward spiral" trajectory. The blue arrows at the bottom indicate the location of grain boundaries. Here solid lines are guides to the eye.}
\label{fig5}
\end{figure}

The second group of experimental results (labeled "test No. 13") were obtained after heating tapes were wrapped around the Nb plate, while the cavity was attached to the vertical stand under vacuum. The Nb plate was baked at $110^\circ$C for 24 h, then the heating tapes were removed, and the test stand was inserted in the cryostat (the solenoid was not attached for these measurements). We observed the low-field $Q_0\simeq 2\times 10^9$ at 2 K and the multipacting-induced quenches at $B_p \simeq 84$ mT as shown in Fig. 6. The residual resistance increased from $\sim 128 \pm 25$ n$\Omega$ to $\sim 175 \pm 78$ n$\Omega$ after baking. The temperature map at $B_p = 60$ mT measured during ramp-down of the rf power is shown in Fig. 7. After the rf power was switched off, the laser beam with the diameter of $0.87$ mm and power of $\simeq 4.4$ W was directed at the locations shown in Fig. 7. For these laser parameters, the peak temperature at the inner Nb surface $\simeq 8.5$ K and the temperature gradient of $\simeq 8$ K/mm were obtained from numerical simulations of $T({\bf r})$ taking into account the temperature dependencies of superconducting and thermal parameters of Nb, and laser absorption coefficient \cite{gigil}. Laser sweeps in the positive x-direction were repeated for multiple initial y-positions to cover an area of $20\times 20$ mm$^2$ around the initial location. The speed of the laser scan is $\simeq 17$ mm/s. Then, LH was repeated with an inward spiral trajectory, beginning at a $36$ mm radius from the center of the plate and ending at the center of the plate. This trajectory was repeated at different speeds, from $\simeq 2$ mm/s and up to $\simeq 32$ mm/s.

%\bigskip
\begin{figure}
\includegraphics[width=\columnwidth]{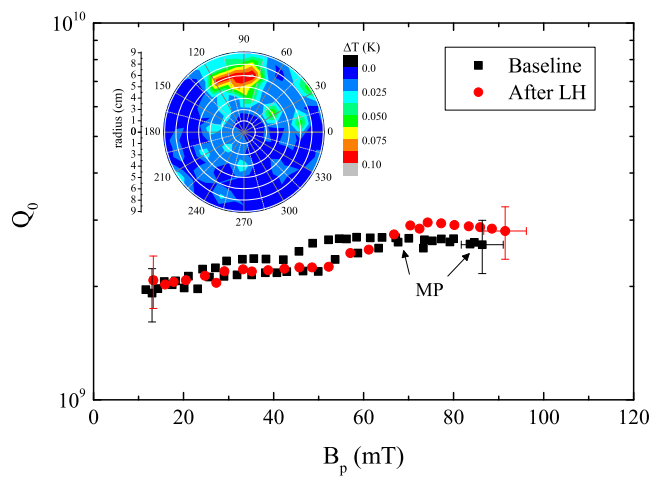}
\caption{(Color) $Q_0(B_p)$ measured at 2 K during test No. 13, before and after laser heating with different laser beam trajectories done at 2 K. A temperature map during MP at the $B_p \simeq 84$ mT is shown in inset.}
\label{fig6}
\end{figure}

\bigskip

\begin{figure}
\includegraphics[width=\columnwidth]{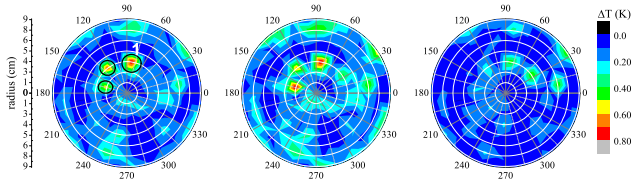}
\caption{Temperature maps at 2 K, $B_p = 60-61$ mT measured during the baseline rf test (left), after LH at 2 K (center) and after LH done at 4.2 K (right) for test No. 13. Red circles show the locations where some of the local LH was done. A 2D map of surface resistance at location "1" obtained by laser scanning microscopy is shown in Fig. 8.}
\label{fig7}
\end{figure}

\bigskip

\begin{figure}
\includegraphics[width=\columnwidth]{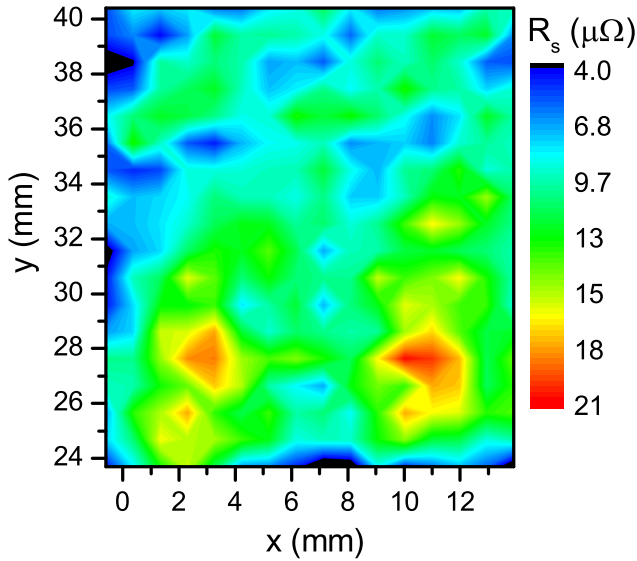}
\caption{(Color) 2D image of the surface resistance  at 2.0 K and $B_p \simeq 13$ mT in the hotspot region labelled as "1" in Fig. 7, obtained by laser scanning microscopy. The laser beam parameters were: 0.87 mm diameter, $\sim$4.4 W power and 10 Hz frequency modulation.}
\label{fig8}
\end{figure}
\begin{figure}
\includegraphics[width=7.5cm]{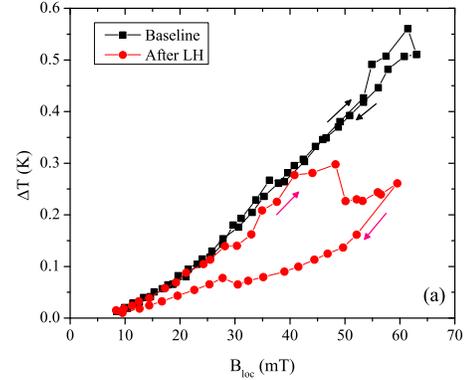}
\includegraphics[width=7.5cm]{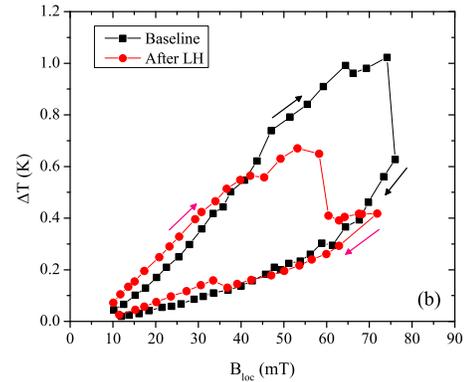}
\caption{$\Delta T(B_{loc})$ measured in test No. 13 at 2 K for two sensors, one at ring 2, at $75^\circ$ (a), the other at ring 3, $169^\circ$ (b) during ramp-up and down of the rf field during the baseline and after LH done at 4.2 K. For the sensors along ring 2, $B_{loc} = 0.726B_p$, while for those along ring 3, $B_{loc} = 0.876B_p$.}
\label{fig9}
\end{figure}

The rf measurements at 2 K following LH revealed MP-induced quench at $B_p \simeq 91$ mT, but no significant changes in either the $Q_0(B_p)$ curves or the temperature maps during ramp-up of the rf field before and after LH (see Figs. 6 and 7 and the temperature map taken during MP in inset of Fig. 6). This experiment was repeated at higher He bath temperature $T_0=4.2$ K by directing the laser beam with the same parameters at the locations shown in Fig. 7. An outward spiral scanning laser beam  trajectory, started at the center of the thermometer and stopped 16 mm away from it was tried at each hotspot location. Then, laser sweeping following an inward spiral trajectory similar to the one done at 2 K, was done at a reduced laser power of $\simeq 1$ W and speed of $\simeq 9$ mm/s. The bath temperature was then lowered back to 2 K by pumping on the He bath and another rf measurement was performed. We observed the low-field $Q_0 \simeq 2\times 10^9$, a multipacting barrier at $B_p \simeq 70$ mT and the multipacting-induced quenches at $B_p \simeq 84$ mT. There was no significant change in the $Q_0(B_p)$ curve as compared to the baseline test or after LH at 2 K, but the temperature maps at $B_p = 61$ mT during ramp-down of the rf field at 4.2 K did change after LH as compared to 2 K. A 2D color map of the local surface resistance of the hotspot region labelled as "1" in Fig. 7 was obtained using the LSM technique \cite{gigil} is shown in Fig. 8. This color map reveals local variations of $R_s$ by the factors $\simeq 4-5$ over spatial scales $\simeq 1-10$ mm. Figure 9 shows the temperature rise as a function of the local amplitude of the rf field ($B_{loc}$) at some thermometer locations, during cycling of the rf power in the baseline as well as in the rf tests after LH at 4.2 K.

The last group of experimental results (labeled "test No. 14") shown here were obtained after the cavity was maintained assembled on the test stand at 300 K and under vacuum, following test No. 13. The baseline rf measurement was done after cooling the cavity to 2 K in the vertical cryostat. We observed the low-field $Q_0 \simeq 2.1\times 10^9$ and strong MP inducing multiple quenches at $B_p \simeq 86 $ mT. He processing did not help us increase the MP barrier. The rf power was cycled up and down twice but no significant change in the $Q_0(B_p)$ curve due to rf cycling was observed. At the same time, the temperature maps at $B_p = 65$ mT measured during the first and the second field ramp up changed as shown in Fig. 10.

%\bigskip
\begin{figure}
\includegraphics[width=\columnwidth]{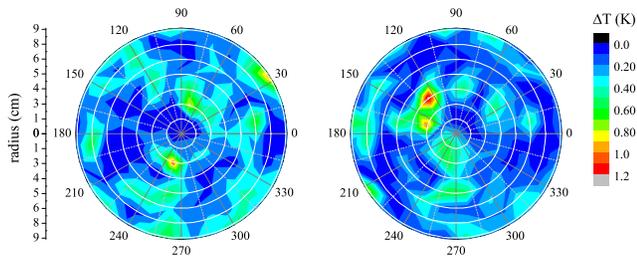}
\caption{Temperature map measured in test No. 14 at 2.0 K, 65 mT during the second field ramp-up in the baseline (left) and during the first field ramp-up after LH following a large "inward spiral" trajectory done at 2.0 K and 4.2 K (right). }
\label{fig10}
\end{figure}
The temperature map at $B_p = 65$ mT during the first field ramp-up after LH is shown in Fig. 10, and the effect of
MP on the temperature map is shown in the inset of Fig. 11. Here LH was done with the laser beam with the power 4.4 W and diameter 0.87 mm, following an inward spiral trajectory, starting at a radius of 38 mm from the center of the cavity plate and ending at the center. The radius decreased by 0.5 mm after each turn and the scanning speed was $\simeq 30$ mm/s. The He bath temperature was then increased to 4.2 K and LH was repeated with the same parameters and trajectory as at 2 K. Finally, the bath temperature was lowered back to 2.0 K by pumping on the He bath and a high-power rf test was performed by again cycling the rf power twice. No significant change in the $Q_0(B_p)$ curve resulted from cycling the rf power and the data for the second ramp-up are shown in Fig. 11. Figure 12 shows an example of $\Delta T(B_{loc})$, measured by the thermometers in ring 2 at $75^\circ$ angle and in ring 3 at $169^\circ$ angle, during the rf cycling, before and after LH.

%\bigskip
\begin{figure}
\includegraphics[width=\columnwidth]{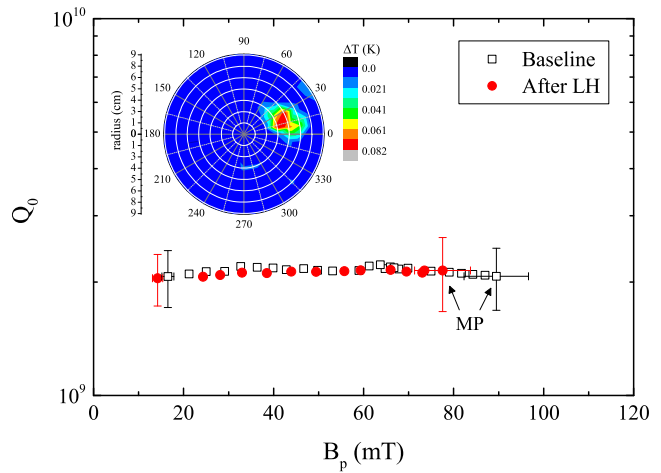}
\caption{$Q_0(B_p)$ measured at 2.0 K in test No. 14, during ramp-down of the rf field before and after laser heating done at both 2 K and 4.2 K. A temperature map during MP at $B_p \simeq 86$ mT is shown in inset.}
\label{fig11}
\end{figure}

We also probed the stability of the superconduting state at various locations at the Nb plate by locally heating it with the laser beam and measuring the rf field at which it can ignite the quench propagation. As example of such measurement is shown in Fig. 13: during test 11c, the laser beam (0.87 mm diameter, 10 W power) was directed at the sensor located on ring 3 at $169^\circ$ and the rf field was increased from zero up to a quench field value $B_{loc} = 64$ mT. In the absence of the laser heating, this area was stable up to higher field $B_{loc} = 75$ mT at which the quench occurred at some other location.

%\bigskip
\begin{figure}
\includegraphics[width=7.5cm]{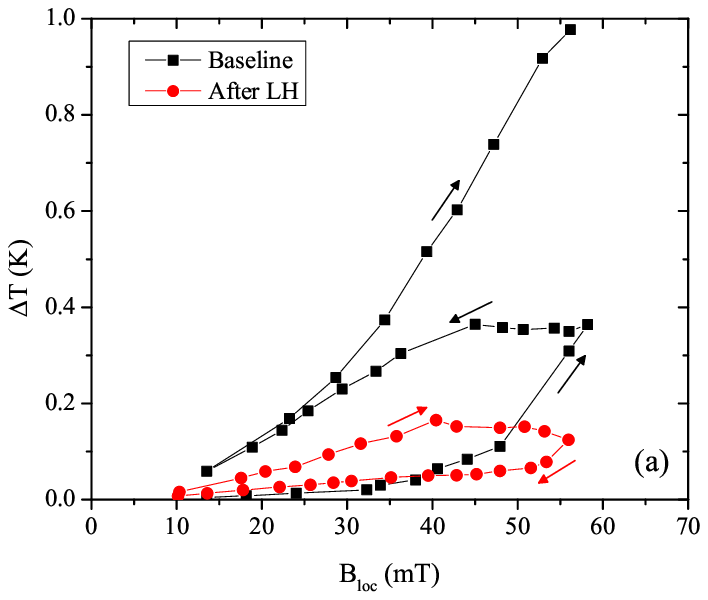}
\includegraphics[width=7.5cm]{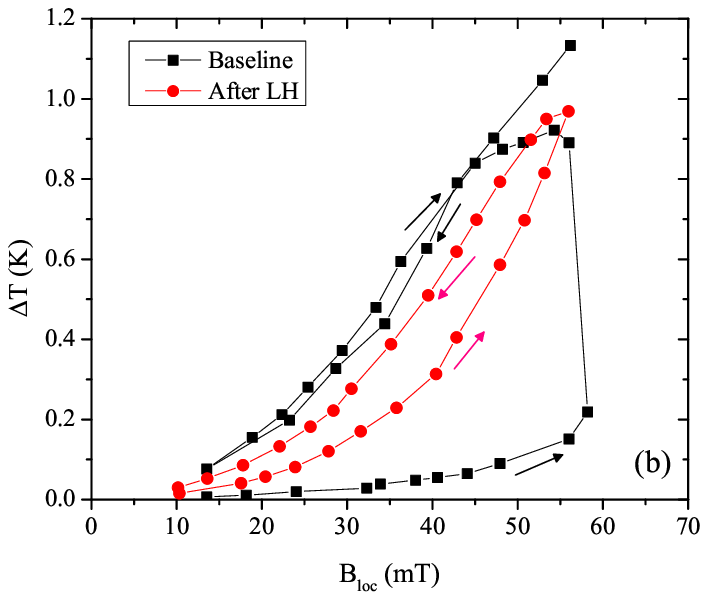}
\caption{$\Delta T(B_{loc})$ measured at 2.0 K for two sensors: in ring 2 at $75^\circ$ (a), in ring 2 at $255^\circ$ (b) while cycling the rf field during the baseline and after LH in test No.14.}
\label{fig12}
\end{figure}

%\bigskip
\begin{figure}
\includegraphics[width=7cm]{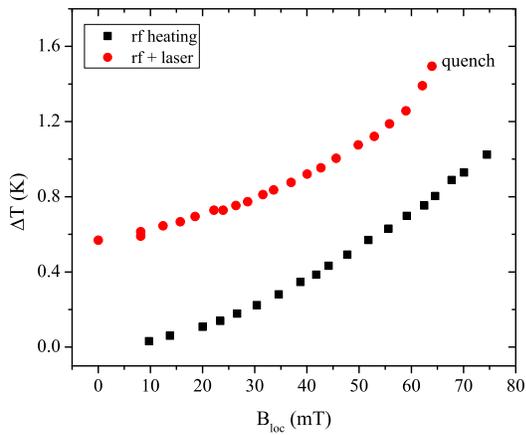}
\caption{$\Delta T(B_{loc})$ measured at 2.0 K for the sensor located in ring 3 at $169^\circ$ with and without additional heating produced by a 10 W laser beam of 0.87 mm in diameter.}
\label{fig13}
\end{figure}

The data presented above show that the scanning laser beam can move and split hotspots on the surface of Nb plate. Moreover, cycling the rf field back-and-forth between the lowest value and the quench field can produce hysteretic changes in the temperature maps, and $\Delta T$ values at particular locations as shown in Figs. 9 and 12. To understand possible mechanisms of such hysteresis, we first notice that redistribution of impurities caused by a low-power laser hotspot moving with the velocities $\sim 1$ cm/s cannot explain it. Indeed, thermally-activated diffusion of such common impurities as C, O or N in Nb by distances $\sim$ few nm typically requires $\sim 10$ hours at $100-200^{\circ}$ C, Ref. [\onlinecite{cav1,diff1,diff2}], so no diffusion redistribution of impurities can occur at liquid helium temperatures. Quantum tunneling of hydrogen interstitials in Nb at low $T$ has been discussed in the literature \cite{H_tunnel}, but tunneling over distances $\sim 1$ cm characteristic of the changes in our temperature maps does not appear plausible.

Another possibility for the hysteretic temperature maps may result from the LH-induced changes in the distribution of multipacting sources at $\simeq 40,\ 70, \ 85$ mT. For the ideal cavity geometry, electron field emission should not occur in the TE$_{011}$ mode, but our 3D numerical simulations have shown that MP may occur in some regions of the flat Nb plate where the TE$_{011}$ mode's cylindrical symmetry is perturbed by the presence of side-ports and coupling antennae \cite{MP_pac}. Multipacting is usually triggered by dust micro particles or nanoscale layers of hydrocarbons which change the secondary emission yield from the Nb surface \cite{cav}. Laser scanning could, in principle, break such absorbed layers, "scraping" the Nb surface off the multipacting sources, but the weak local overheating $\lesssim 10$ K due to absorption of $2.3$ eV photons produced by our low-power scanning laser is not sufficient to cause any chemical changes of adsorbates, which typically require much higher laser powers and photon energies in the ultraviolet spectrum \cite{lasersurf}. In any case, had LH somehow deactivated the MP sources on the Nb surface, the hysteresis in the temperature maps would have disappeared after the first laser scan. However, we observed consistent hysteretic changes in the temperature maps after repeatedly cycling the rf field (although $\Delta T(B_{loc})$ in some locations did not always follow the previous hysteretic loop).

Based on the above experimental data, we conclude that the hysteretic temperature maps can be explained by the presence of
trapped vortex bundles - the only objects in a superconductor which can be re-distributed over the surface by a low-power scanning laser beam producing local overheating of only a few degrees. For instance, the local laser overheating by $\simeq 7$ K can turn the top of the hot region in Fig. 1 in the normal state, depinning the tips of all vortices in this area. As the laser beam moves, the tips of the vortex lines get redistributed and stuck in other configurations in the random array of pinning defects at the surface. The fact that some of the hotspots do not disappear but become either weaker or stronger is consistent with the redistribution of trapped vortex lines shown in Fig. 1 which can also explain the hysteretic behavior of $\Delta T(B_p)$ at upon cycling the rf power up to the quench field. The field dependence of $\Delta T$ at hotspots can be described as $\Delta T \propto B_p^n$ with $n$ ranging between 1.5 and 3, as shown in Fig. 12. From the thermal map and scanning laser data, we can infer an information about local dissipation sources, particularly trapped vortex bundles, as will be shown below.

\section{DISSIPATION DUE TO TRAPPED VORTICES}

\subsection{Dynamic equations}

Based on the qualitative picture shown in Fig. 1, we calculate the power $P$ dissipated by a perpendicular vortex segment pinned by a defect spaced by $\ell $
from the surface, as shown in Fig. 14. In this work the effect of thermal fluctuations of vortices on $R_s$ at $T\ll T_c$ (see, e.g.,  Ref. \onlinecite{cc}) is neglected. The dynamic equation for the vortex in a weak rf field $H(z,t)=H_p\exp(-z/\lambda + i\omega t)$ parallel to the surface
is given by
\begin{equation}
\eta \dot{u}=\hat{\varepsilon}u'' +F\exp (-z/\lambda +i\omega t),
\label{deq}
\end{equation}
where $z$ is the coordinate perpendicular to the surface, $u(z,t)$ is the
vortex displacement parallel to the surface, $F=\phi _{0}H_p/\lambda$
is the amplitude of the rf driving force, $\lambda$ is the London penetration depth in the $ab$ plane, $\eta = \phi_0 B_{c2}/\rho_n$ is the viscous drag coefficient, where $B_{c2}=\phi_0/2\pi\xi^2$ is the upper critical field, $\rho_n$ is the normal state resistivity, and $\xi$ is the coherence length. The operator $\hat{\varepsilon}$ describes the dispersive vortex line tension in a uniaxial superconductor with the c-axis perpendicular to the surface. The Fourier transform of $\hat{\varepsilon}$ is \cite{ehb,geshk}
\begin{equation}
\varepsilon (k)=\frac{\varepsilon _{0}}{2\Gamma^2}\ln \frac{\kappa_{GL} ^{2}\Gamma^2}{1+\lambda
^{2}k^{2}}+\frac{\varepsilon _{0}}{2\lambda ^{2}k^{2}}\ln (1+\lambda
^{2}k^{2}),
\label{eps}
\end{equation}
where $\varepsilon _{0}=\phi _{0}^{2}/4\pi \mu _{0}\lambda ^{2}$,
$\kappa_{GL} = \lambda/\xi$ is the GL parameter, and $\Gamma=\lambda_c/\lambda$ is the
anisotropy parameter. We calculate $P(\omega,\ell)$ for a vortex line trapped by sparse pinning centers
(for example, oxide nanoprecipitates), following our previous calculations of $R_s$ for pinned vortex segments parallel to
the surface \cite{gurci}. This approach takes into account all bending modes of a vibrating vortex segment (the case of a
vortex trapped by a columnar defect perpendicular to the surface was considered in Ref. [\onlinecite{lemp}]), unlike
theories of microwave response \cite{gr,rabin,cc,ehb} using a phenomenological Labusch pinning spring constant for a vortex regarded as a particle rather
than as an elastic string.

\begin{figure}[ht]
\includegraphics[width=7cm,angle=0]{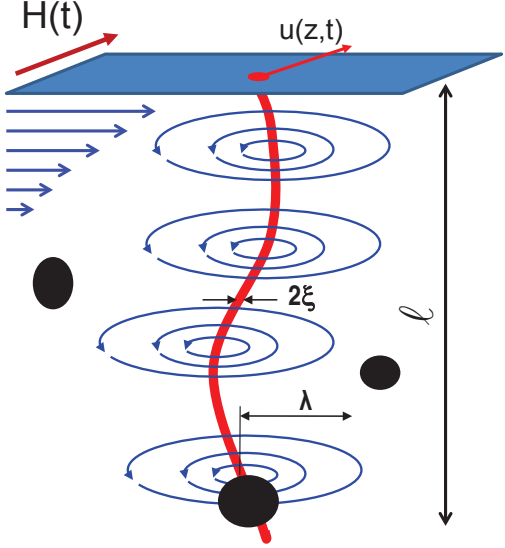}
    \caption{Oscillating vortex segment pinned by a defect spaced by $\ell$ from the surface in the presence of rf
    Meissner currents.}
    \label{fig14}
\end{figure}

For strong core pinning, the boundary conditions to Eq. (\ref{deq}) are that one end of the vortex is perpendicular to the
surface \cite{ehb}, and the other end is fixed by the pin:
\begin{equation}
u' (0)=0,\qquad u(\ell )=0 \label{bc}
\end{equation}
The solution of Eq. (\ref{deq}) which satisfies Eq. (\ref{bc}) is
\begin{equation}
u(x,t)=\sum_{n=0}^\infty A_{n}\cos (k_{n}z)e^{i\omega t},\quad k_{n}=\frac{\pi }{%
\ell }\left( n+\frac{1}{2}\right).
\label{u}
\end{equation}
Substituting Eq. (\ref{u}) into Eq. (\ref{deq}) multiplied by $\cos k_{n}z$ and
integrating over $z$ from $0$ to $\ell$, yields
\begin{gather}
A_{n}=\frac{FI_{n}}{i\omega \eta +k_{n}^{2}\varepsilon (k_{n})},
\label{A} \\
I_{n}=\frac{\pi (2n+1)(-1)^{n}e^{-a}+2a}{a^{2}+\pi ^{2}(n+1/2)^{2}},
\label{In}
\end{gather}
where $a=\ell /\lambda $.  Notice that Eq. (\ref{u}) is not a complete solution if $\ell <\lambda$ so that the rf Meissner currents
can reach trapped segments of the vortex behind the first pin at $z>\ell$. We first consider the case of $\ell > \lambda$ for which only the nearest to the
surface vortex segment is excited, and then address the case of $\ell < \lambda$ (particularly relevant to thin films) in subsection C.

If the spatial dispersion of $\varepsilon(k_z)$ can be neglected, the solution of Eq. (\ref{deq}) becomes
\begin{gather}
u(z,t)=\frac{H_p\phi_0\lambda e^{i\omega t}}{i\omega\eta\lambda^2-\varepsilon}\biggl[e^{-z/\lambda}+\frac{1}{q_\omega\lambda}\sinh(q_\omega z) \nonumber \\
-\frac{\cosh(q_\omega z)}{\cosh(q_\omega\ell)}\biggl(e^{-\ell/\lambda}+\frac{1}{q_\omega\lambda}\sinh(q_\omega\ell)\biggr)\biggr],
\label{uloc}
\end{gather}
where $q_\omega = (i\eta\omega/\varepsilon)^{1/2}$, and $\varepsilon=\varepsilon(k_z\rightarrow 0)$. In isotropic superconductors, $\varepsilon = \phi_0H_{c1}$, where
$H_{c1}$ is the lower critical field. The anisotropy affects $\varepsilon$ as follows:
\begin{equation}
\varepsilon = \frac{\phi_0^2g}{4\pi\mu_0\lambda^2}, \qquad g = \frac{1}{\Gamma^2}\ln(\Gamma\kappa_{GL})+\frac{1}{2}.
\label{epso}
\end{equation}

Equation (\ref{uloc}) shows that, the amplitude of driven rf oscillations
of a long vortex segment is maximum at the surface and decays along $z$ over the length which depends on $\omega$. At low frequencies,
$\omega \lesssim \omega_l$ where $\ell q_\omega =1$, the surface
rf Meissner current at $0<z<\lambda$ causes rocking of the whole vortex segment of length $\ell \gg \lambda$.
At $\omega_l \lesssim\omega \lesssim\omega_\lambda$ where $\lambda q_\omega = 1$,
the rf oscillations of the vortex are mostly localized in the surface layer of thickness of $1/q_\omega$ smaller than $\ell$ but larger
than $\lambda$. At higher frequencies $\omega \gtrsim \omega_\lambda$, only a short tip $\sim \lambda$ of the vortex is driven by the rf currents.
Here $\omega_\lambda$ and $\omega_l$ are given by:
\begin{equation}
\omega_\lambda=\frac{g\rho_n\xi^2}{2\mu_0\lambda^4}, \qquad \omega_l=\frac{g\rho_n\xi^2}{2\mu_0\lambda^2\ell^2}.
\label{om}
\end{equation}
The rf power $P(\omega)$ behaves quite differently in these frequency domains
$0<\omega <\omega_l$, $\omega_l < \omega < \omega_\lambda$, and $\omega > \omega_\lambda$,
and becomes independent of pinning at $\omega>\omega_l$.

\subsection{Dissipated power in a semi-infinite superconductor}

We first consider the case of $\ell\gg\lambda$ for which the Meissner current density is negligible at the pin and only one vortex segment at $0<z<\ell$ is excited. Using Eqs. (\ref{u})-(\ref{In}), we obtain the mean power $P=-(\omega F/2)\mbox{Im}\int_{0}^{\ell }u(z,\omega)\exp(-x/\lambda )dx$ dissipated by the vibrating vortex segment, as shown in Appendix A:
\begin{equation}
P=\frac{H_p^2\omega^2}{4\lambda^2}\sum_{n=0}^\infty\frac{\eta\phi_0^2\ell I_{n}^2}{\omega ^{2}\eta ^{2}+k_{n}^{4}\varepsilon ^{2}(k_{n})}.
\label{Qg}
\end{equation}
This equation enables one to calculate $P(\omega,\ell)$, taking into account both the nonlocality of the vortex line tension and the
uniaxial anisotropy essential in layered materials like high-$T_c$ cuprates. An example of such calculation
for $\ell=4\lambda$, $\lambda/\xi=100$ and different anisotropy parameters $\Gamma$ is shown in Fig. 15. Here $P(\omega)$ first increases quadratically with $\omega$,  then goes like $P\propto \omega^{1/2}$ at $\omega_l\ll \omega \ll \omega_\lambda$, and saturates at higher $\omega$.  The behavior of $P(\omega)$ calculated from Eq. (\ref{Qg}) taking into account all bending vortex modes is more complicated than $P\propto \omega^2/(\omega_p^2+\omega^2)$ of the Gittleman and Rosenblum model \cite{gm} which contains only one phenomenological depinning frequency $\omega_p$.

%\bigskip
\begin{figure}[ht]
\includegraphics[width=6.8cm,angle=0]{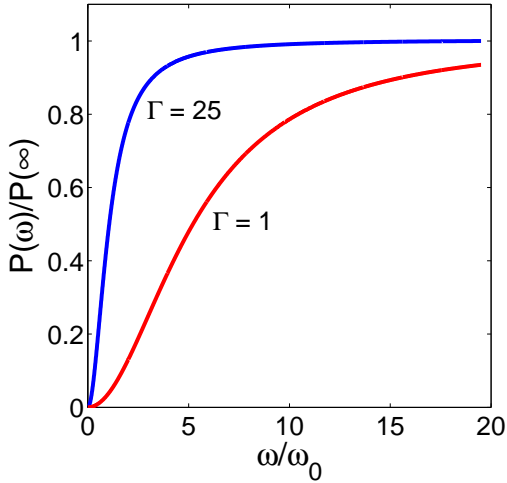}
    \caption{The frequency dependence of $P(\omega)$ calculated from Eq. (\ref{Qg}) for $\ell=4\lambda$, $\kappa_{GL}=100$ and different anisotropy
    parameters $\Gamma=25$ and $\Gamma=1$. Here $P(\omega)$ is normalized to the maximum $P(\omega \rightarrow \infty)$, and $\omega$ is normalized to
    $\omega_0 = \epsilon_0/2\eta\lambda^2$.}
    \label{fig15}
\end{figure}

The anisotropy increases $P(\omega)$ at low and intermediate frequencies, but does not affect $P$ at $\omega\gg\omega_\lambda$. This large anisotropy parameter $\Gamma$ reduces the vortex line tension $\epsilon(k)$ in Eq. (\ref{eps}), particularly at $k\lambda>1$. The anisotropy thus reduces the frequencies at which the term $k_n^4\epsilon^2(k_n)$ in the denominator of Eq. (\ref{Qg}) becomes negligible so that $P(\omega)$ levels off as the factors $\propto \omega^2$ cancel out.

For long vortex segments, $\varepsilon (k_z)$ can be taken in the local limit at $k_z=0$
since the main contribution to the sum in Eq. (\ref{Qg}) comes from $k_{n}\lambda <1$. In this case
the summation can be done exactly, or one can directly use Eq. (\ref{uloc}) to calculate $P$, as shown in Appendix A. The result
can be presented in two equivalent forms:
\begin{gather}
P=\frac{H_p^2\phi_0^2\chi^2}{4\eta\lambda}\biggl[\frac{5+\chi^2}{(1+\chi^2)^2}-\frac{2}{\chi^{3/2}}\mbox{Im}\frac{\tanh\sqrt{i\nu}}{\sqrt{i}(1-i\chi)^2}\biggr],
\label{Q1}\\
\chi=\omega\eta\lambda^{2}/\varepsilon, \qquad \nu=\omega\eta\ell^2/\varepsilon.
\label{nu}
\end{gather}
Separating here the imaginary part yields
\begin{gather}
P=\frac{H_p^{2}\phi _{0}^{2}\chi^{2}}{2\eta\lambda
(1+\chi^{2})^{2}}\biggl[\frac{\chi^2}{2}+ \frac{5}{2} + \nonumber \\
\frac{
(1-2\chi-\chi^2)\sinh\sqrt{2\nu}-(1+2\chi-\chi^{2})\sin\sqrt{2\nu}}{
\sqrt{2}\chi^{3/2}(\cosh\sqrt{2\nu}+\cos\sqrt{2\nu})}\biggr].
\label{Q2}
\end{gather}
Here $\chi=\omega/\omega_\lambda$ and $\nu=\omega/\omega_l$ are dimensionless frequencies where
$\omega_\lambda = \varepsilon/\eta\lambda^2$ and $\omega_l = \varepsilon/\eta\ell^2$ are defined by Eq. (\ref{om}).
Taking $\xi/\lambda = 1$, $\rho_n = 10^{-9}$ $\Omega$m and $\lambda = 40$ nm for clean Nb at $T\ll T_c$,
we obtain $\omega_\lambda \simeq 2.5\times 10^{11}$ Hz, about one tenth of the gap frequency
$\omega_\Delta\sim 1.76k_BT_c/\hbar \simeq 2.3\times 10^{12}$ Hz above which the rf field generates quasi-particles at
$T\ll T_c$. The frequency $\omega_\lambda (T)\simeq \omega_\lambda (0)(1-T^2/T_c^2)$ decreases as
$T$ decreases, vanishing at $T_c$. By contrast, $\omega_l(T)$ is
nearly temperature-independent and remains finite at $T_c$. Thus, for large vortex segments, $\ell\gg \lambda$, we have
$\omega_\lambda (T) \gg \omega_l(T)$ at $T\ll T_c$, but $\omega_\lambda (T) \ll \omega_l(T)$ if $T$ is close to $T_c$.

For $\chi=\omega/\omega_\lambda\ll 1$, Eq. (\ref{Q2}) simplifies to:
\begin{equation}
P=\frac{H_p^{2}\phi _{0}^{2}(\sinh\sqrt{2\nu}-\sin\sqrt{2\nu})\sqrt{\nu}}{2^{3/2}\eta\ell(\cosh\sqrt{2\nu}+\cos\sqrt{2\nu})}
\label{Qm}
\end{equation}
At low frequencies, $\nu\ll 1$, Eq. (\ref{Qm}) takes the form
\begin{equation}
P=4\pi H_p^2\lambda ^4\mu_0^2\ell^3\omega ^2/3\rho_n g^2\xi^2 ,\qquad
\omega\ll \omega_l
\label{P1}
\end{equation}
Here $P$ decreases as the mean spacing $\ell$ between pins decreases, unlike the region $\omega\gg\omega_l$ in which
$P$ becomes independent of pinning. Indeed, at intermediate frequencies, $\nu\gg 1$, Eq. (\ref{Qm}) yields:
\begin{equation}
P=\pi H_p^2\lambda\xi(\mu_0\rho_n\omega/g)^{1/2}, \qquad \omega_l \ll \omega \ll \omega_\lambda
\label{P2}
\end{equation}
At high frequencies, $\chi=\omega\eta\lambda^2/\varepsilon\gg 1$, $P$ becomes independent of
$\omega$, $\ell$, and the vortex line tension:
\begin{equation}
P=\pi H_p^2\xi^2\rho_n/2\lambda,\qquad \omega\gg \omega_\lambda
\label{P3}
\end{equation}

It is instructive to trace the effect of nonmagnetic impurities on $P(\omega)$ in different frequency regions, taking into account
the dependencies of $\rho_n\propto l_i^{-1}$, $\xi\propto l_i^{1/2}$, and $\lambda\propto l_i^{-1/2}$ on the mean free path $l_i$ in the dirty limit.
Then Eq. (\ref{P1}) shows that $P\propto l_i^{-2}$ at $\omega\ll\omega_l$ increases strongly as $l_i$ decreases. At intermediate frequencies,
$\omega_l\lesssim\omega\lesssim\omega_\lambda$, the power $P\propto l_i^{-1/2}$ keeps increasing upon decreasing $l_i$ but much weaker than at $\omega\ll \omega_l$. This trend reverses at high frequencies $\omega\gg \omega_\lambda$ for which $P\propto l_i^{1/2}$ decreases as the surface gets dirtier.

We estimate $P$ in Nb at frequencies $\omega\gg\omega_l$ for which $P(\omega)$ is independent of $\ell$. Taking $\rho_n=10^{-9}$ $\Omega$ m, $\xi=\lambda = 40$ nm for clean Nb, we obtain from Eq. (\ref{P2}) that $P\simeq 0.13\ \mu$W for $B_p = 100$ mT and $\omega/2\pi = 2$ GHz. This estimate corresponds to the intermediate frequencies $\omega_l\lesssim\omega\lesssim\omega_\lambda$ relevant to our experiment. For the same materials parameters, the high-frequency limit of $P$ defined by Eq. (\ref{P3}) yields $P\simeq 0.4\ \mu$W.

\subsection{Vortices in a thin film}

The results of the previous subsection can be used to calculate $P$ for a perpendicular vortex in a thin film with $d\ll \lambda$ relevant to the rf dissipation in thin film multilayers in accelerator cavities \cite{ml} and the nanoscale thin film structures in superconducting quibits and photon detectors. We consider a vortex pinned by a single defect spaced by $\ell$ from the film surface, as shown in Fig. 16, for which $P$ essentially depends on the way by which the magnetic field is applied. In the first case the rf filed is applied to one side of a thin film screen so that the Meissner current density $J\approx H_p/\lambda$ is nearly uniform over the film thickness. The second case corresponds to a film in a uniform parallel rf field, for which $J(z)=(H_p/\lambda)\sinh(z/\lambda)/\cosh(d/2\lambda)$ changes sign in the middle of the film at $z=0$.

\bigskip
\begin{figure}[ht]
\includegraphics[width=6.8cm,angle=0]{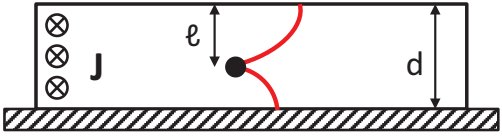}
    \caption{A vortex pinned by a single defect spaced by $\ell$ from the surface of a thin film screen with $d\ll\lambda$. Under the action of uniform Meissner currents flowing perpendicular to the plane of the picture, two vortex segments bow out in such a way that their tips remain perpendicular to the film surface. }
    \label{fig16}
\end{figure}

In a thin film screen the rf power dissipated by two vortex segments of length $\ell$ and $d-\ell$ is given by Eq. (\ref{Qg}) with $I_n \to 4(-1)^n/\pi(2n+1)$ being the limit of Eq. (\ref{In}) at $a=\ell/\lambda \ll 1$:
\begin{eqnarray}
P=\frac{4\eta H_p^2\phi_0^2\omega^2}{\pi^2\lambda^2}\sum_{n=0}^\infty\frac{1}{(2n+1)^2}\times \nonumber \\
\left[\frac{\ell}{\omega ^{2}\eta ^{2}+k_{n}^{4}\varepsilon ^{2}(k_{n})}+\frac{d-\ell}{\omega ^{2}\eta ^{2}+q_{n}^{4}\varepsilon ^{2}(q_{n})}\right],
\label{Qf}
\end{eqnarray}
where $q_n=\pi(n+1/2)/(d-\ell)$. Equation (\ref{Qf}) simplifies at low frequencies $\omega \ll \pi^2\varepsilon(\pi/d)/\eta d^2$ which can extend to the THz region
for thin films with $d\ll \lambda$. In this case the terms of the sum in Eq. (\ref{Qf}) decrease very rapidly with $n$ so to the accuracy of
better than $2\%$, we can retain only terms with $n=0$:
\begin{equation}
P\simeq \frac{64\eta H_p^2\phi_0^2\omega^2}{\pi^6\lambda^2}
\left\{\frac{\ell^5}{\varepsilon ^{2}[\pi/2\ell]}+\frac{(d-\ell)^5}{\varepsilon ^{2}[\pi/2(d-\ell)]}\right\}.
\label{Qff}
\end{equation}
Here $P$ is proportional to $\omega^2$ and decreases rapidly as the film thickness decreases. Strong anisotropy $\Gamma\gg 1$ and
elastic nonlocality of the vortex line tension at $\pi\lambda/d \gg 1$ increase $P$ in Eq. (\ref{Qff}) because both effects reduce
the vortex line tension $\varepsilon(k)$ defined by Eq. (\ref{eps}).

The vortex in the screen gets depinned as the upper and lower segments of the bowed vortex
become parallel and reconnect at the pin if $H_p>H_{pin}$. We evaluate $H_{pin}$ for the symmetric case of $\ell=d/2$, using the expression \cite{brandt} for
the pin breaking current density, $J_c = (\phi_0/4\pi\mu_0\lambda^2\Gamma\ell)\ln(2\ell\Gamma/\xi)$. Hence,
\begin{equation}
B_{pin} \simeq \frac{\phi_0}{4\pi\lambda_c\ell}\ln\frac{2\ell}{\xi_c},
\label{hpin}
\end{equation}
where $\lambda_c$ and $\xi_c$ are the penetration depth and coherence length along the c-axis in a uniaxial superconductor. The depinning field $B_{pin}\sim B_c\xi/\Gamma\ell$ can be much smaller than the thermodynamic critical field $B_c = \phi_0/2\sqrt{2}\pi\lambda\xi$, particularly for highly anisotropic materials.

Now we turn to a film in a parallel field, limiting ourselves to the symmetric case $\ell = d/2$. Here $P$ is given by Eq. (\ref{Qg}) in which the form factor $I_n$ is replaced with $\tilde{I}_n=(4/d)\int_0^{d/2} \sinh(z/\lambda)\sin(k_nz)dz$, and $k_n=\pi(2n+1)/d$ (see Appendix A):
$$
\tilde{I}_n\simeq \frac{4(-1)^n d}{\pi^2(2n+1)^2\lambda}, \qquad \ell\ll\lambda.
$$
Therefore,
\begin{gather}
P=\frac{4H_p^2\omega^2}{\pi^4\lambda^4}\sum_{n=0}^\infty\frac{\eta\phi_0^2d^3}{(2n+1)^4[\omega ^{2}\eta ^{2}+k_{n}^{4}\varepsilon^{2}(k_{n})]}
\nonumber  \\
\simeq \frac{4\eta H_p^2\phi_0^2d^7\omega^2}{\pi^8\lambda^4\varepsilon^2(\pi/d)},
\label{Pfilm}
\end{gather}
where the terms $\propto (2n+1)^{-8}$ with $n\neq 0$ in the rapidly converging sum were neglected. Comparing Eq. (\ref{Qff})
with Eq. (\ref{Pfilm}) shows that $P$ in a film in a parallel uniform rf magnetic field is reduced by the factor $\simeq (d/\pi\lambda)^2\ll 1$ as compared to a film screen. This is because the screening current density $J\simeq H_p z/\lambda$ which changes sign
at the center of the film, produces much weaker rf drive than the uniform Meissner current in a screen.

\subsection{Residual resistance due to vortices}

Trapped vortices contribute to the residual surface resistance $R_i$ which defines
the dissipated power $P=R_iH_p^2/2$ per unit surface area. Assuming that vortices with the average density $B_0/\phi$ appear due to
a dc magnetic field $B_0$, we obtain $R_i$ using Eq. (\ref{Qg}):
\begin{equation}
R_i=\frac{\eta B_0\phi_0\omega^2}{2\lambda^2}\int_0^\infty\sum_{n=0}^\infty\frac{I_{n}^2 G(\ell)\ell d\ell}{\omega ^{2}\eta ^{2}+k_{n}^{4}\varepsilon ^{2}(k_{n})}.
\label{Ri}
\end{equation}
Here $R_i$ is averaged over a statistical distribution of noninteracting vortex segments with the distribution function
$G(\ell)$ normalized by the condition $\int_0^\infty G(\ell)d\ell =1$.

The frequency dependence of $R_i(\omega)$ is similar to that is shown in Fig. 15. We evaluate $R_i$ at $\omega_l\ll\omega\ll\omega_\lambda$ where $P$ is independent of $\ell$. Then Eq. (\ref{P2}) yields
\begin{equation}
R_i = \frac{B_0}{B_c}\left(\frac{\mu_0\rho_n\omega}{2g}\right)^{1/2}.
\label{ri}
\end{equation}
Let us estimate the magnitude of $B_0$ which gives rise to the observed $R_i=5$ n$\Omega$ in Nb at $f=2$ GHz \cite{cav,gigiJAP}. Taking $\rho_n=10^{-9}$ $\Omega$m,
$B_c=0.2$ T, and $2g=1$ from Eq. (\ref{epso}), we obtain that $R_i=5$ n$\Omega$ can result from the residual field $B_0\simeq 2.5\ \mu$T
much smaller than the Earth field.

In the literature $R_i$ is sometimes evaluated using the  formula
$R_i^H=(B_0/2B_{c2})(\mu_0\rho_n\omega/2)^{1/2}$, assuming that $R_i^H$ is just the surface resistance in the normal state
$(\mu_0\rho_n\omega/2)^{1/2}$ times the volume fraction of vortex cores $B_0/2B_{c2}$ regarded as fixed normal tubes of radius $\xi$
(see, e.g., Ref. \onlinecite{cav}). The so-defined $R_i^H$ ignores the oscillations of vortices under the rf field and
underestimates $R_i$ as compared to Eq. (\ref{ri}) derived for $\omega_l\ll\omega\ll\omega_\lambda$ by the factor
$\simeq 4$ for Nb but $\simeq 40$ for Nb$_3$Sn. Moreover, $R_i^H\propto \rho_n^{-1/2}$
decreases as the mean free path decreases, while $R_i \propto \rho_n^{1/2}$ in Eq. (\ref{ri}) increases as the material gets dirtier since $B_c$ is not affected by nonmagnetic impurities \cite{book}. Equation (\ref{ri}) suggests that $R_i$ may increase in superconductors with higher $\rho_n$, such as Nb$_3$Sn, high-$T_c$ cuprates or semi-metallic Fe-pnictides \cite{as}. For Nb$_3$Sn with $B_c\simeq 540$ mT and $\rho_n\simeq 0.2\ \mu\Omega$ m, we obtain $R_i\simeq 1\ \mu\Omega$ at $f=2$ GHz and $B_0=40\ \mu$T.

Although the thin film geometry facilitates trapping perpendicular vortices, pinning can reduce $R_i$. Indeed, for a thin film screen,
Eq. (\ref{Qff}) with $\ell = d/2$ yields:
\begin{equation}
R_i\simeq \frac{8\phi_0^2B_{c2}B_0 d^5\omega^2}{\pi^6\rho_n\lambda^2\varepsilon^2(\pi/d)}, \qquad d\ll\lambda.
\label{ris}
\end{equation}
For a thin film in a parallel rf magnetic field, $R_i$ readily follows from Eq. (\ref{Pfilm}):
\begin{equation}
R_i\simeq \frac{8\phi_0^2B_{c2}B_0 d^7\omega^2}{\pi^8\rho_n\lambda^4\varepsilon^2(\pi/d)}, \qquad d\ll\lambda.
\label{rif}
\end{equation}
In the dirty limit the ratio $B_{c2}/\rho_n$ is independent of the mean free path $l_i$, while according to  Eq. (\ref{eps}), the products $\lambda^2\varepsilon^2(\pi/d)$ in Eq. (\ref{ris}) and $\lambda^4\varepsilon^2(\pi/d)$ in Eq. (\ref{rif}) decrease as $l_i$ decreases. In both cases $R_i$ increases as the concentration of nonmagnetic impurities increases, although this increase is much slower for a film in a uniform field. Reduction of $R_i$ by denser pinning nanostructure (shorted lengths $\ell$ of vortex segments) is consistent with low $R_i\sim 2-5$ n$\Omega$ observed on Nb films \cite{nbfilm}, Nb$_3$Sn films \cite{nb3sn1,nb3sn2} at 1 GHz, and a significant decrease of $R_i$ due to incorporation of BaZrO$_3$ oxide nanoparticles in YBa$_2$Cu$_3$O$_{7-x}$ films \cite{ybcofilm}.

\section{Temperature distributions in vortex hotspots}

\subsection{Uniform rf heating}

The rf fields can make the Meissner state unstable due to the positive feedback between the exponential temperature dependence of the rf power of $R_s(T)H_p^2/2$ and heat transfer. Here we outline a thermal breakdown model \cite{agac,ag} to introduce the parameters which will be used for the analysis of vortex hotspots. We consider a slab of thickness $d\gg \lambda$ exposed to the rf field at one side $(z=0)$ and cooled at the other $(z=d)$, so that the rf power released in a narrow layer at $z=0$ is balanced by heat diffusion across the slab, as shown in Fig. 17a. Steady-state distribution $T(z)$ and the surface temperatures $T_m=T(0)$ and $T_s=T(d)$ are determined by the boundary conditions: $\kappa T' = - R_sH_p^2/2$ at $z = +0$ and $\kappa T' + \alpha_K(T_s,T_0)(T_s - T_0) = 0$ at $z = d$, where $\kappa$ is the thermal conductivity and $\alpha_K(T_s,T_0)$ is the Kapitza thermal conductance at the cooled surface.

%\bigskip
\begin{figure}
\centerline{\psfig{file=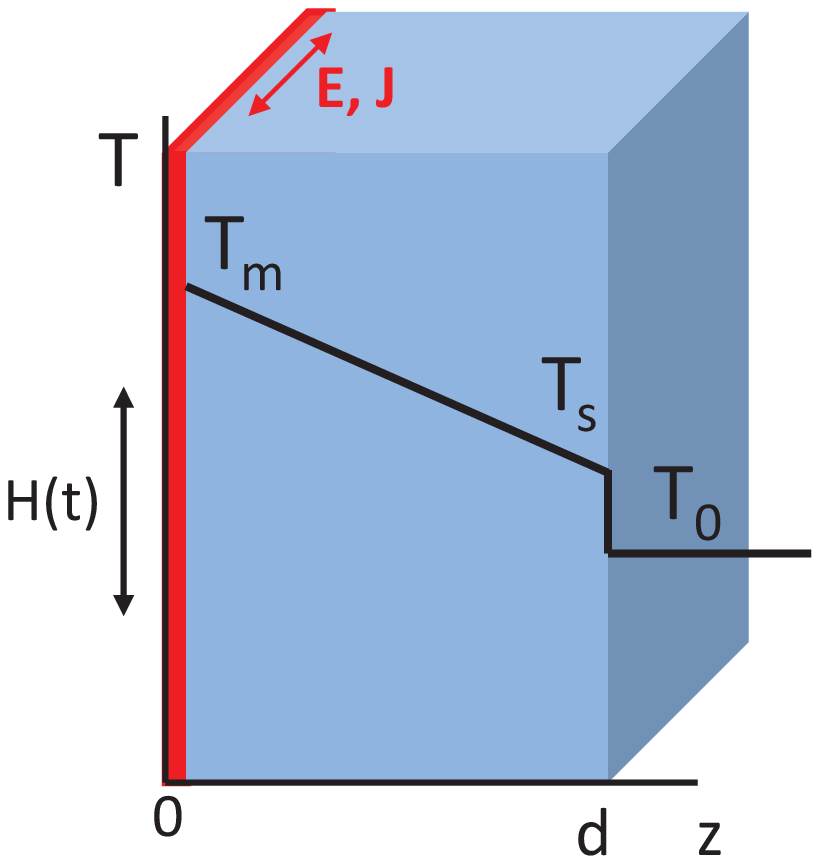,width=4.8cm}}
\centerline{\psfig{file=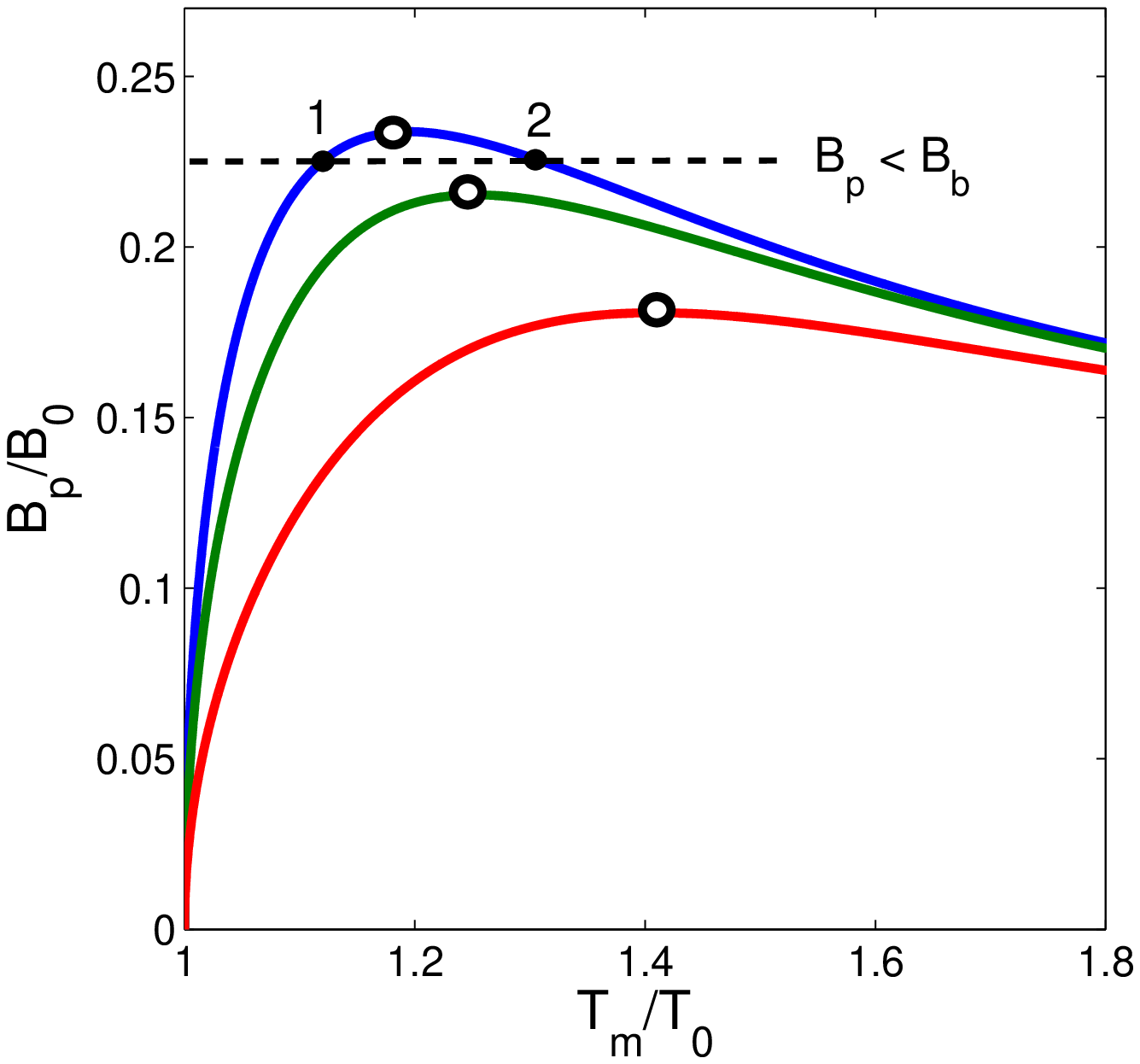,width=5.2cm}}
\caption{(a) Temperature distribution across the sample where red shows
the thin $\sim\lambda\ll d$ layer of rf dissipation. The temperature jump $T_s - T_0$ at $z=d$
is due to the Kapitza thermal resistance between the superconductor and the coolant/sunstrate. (b) graphic solution of
Eq. (\ref{bp}) for different ratios of $R_i/R_{BCS}(T_0) = 0.1; 1; 5$ (from top to bottom). The points 1 and 2 correspond to the
stable and unstable solutions, respectively, and the empty circles show the breakdown fields. Here $B_0^2=2\mu_0^2T_0\kappa\alpha_K/(\kappa+d\alpha_K)R_{BCS}(T_0)$. }
\label{fig17}
\end{figure}

The rf heating results in a field dependence of the surface resistance $R_s \propto \exp[-\Delta/T_m(H_p)]$ yet the temperature raise $T_m - T_0\simeq k_BT_0^2/\Delta \ll T_0$ remains small even at the breakdown field \cite{agac,ag}. Then the temperature dependencies of $\kappa(T)$ and $\alpha_K(T,T_0)$ \cite{cryogen,kapitza} can be neglected as compared to $R_s(T)$, and the thermal flux conservation yields $(T_m-T_s)\kappa/d = (T_s-T_0)\alpha_K$. Here $T_m(H_p)$ is determined self-consistently by the heat balance equation $H_p^2R_s(T_m)/2=\alpha_K(T_s-T_0)$ which is convenient to present in terms of $B_p=\mu_0H_p$ as a function of $T_m$:
\begin{equation}
B_p^2=\frac{2\mu_0^2\kappa\alpha_K T_m(T_m-T_0)}{[A\omega^2\exp(-\Delta/k_BT_m)+T_mR_i](\kappa+\alpha_K d)},
\label{bp}
\end{equation}
where the thermal impedance $\alpha_K\kappa/(\kappa+\alpha_Kd)$ is determined by heat diffusion across the slab and the Kapitza thermal conductance at the surface.

The behavior of $T_m(B_p)$ can be understood from the graphical solution of Eq. (\ref{bp}) shown in Fig. 17b. Here point 1
corresponds to a stable state for which $T_m(B_p)$ increases as $B_p$ increases, while point 2 corresponds to an unstable equilibrium.
The points 1 and 2 merge as $B_p(T_m)$ reaches maximum at $T_m=T_b$, defining the breakdown field $B_b=B_p(T_b)$ above which thermal runaway occurs. Here $B_b$ can be obtained from Eq. (\ref{bp}) and $\partial B_p/\partial T_m=0$. The breakdown occurs at small overheating $\theta = (T_m-T_0)/T_0\ll 1$ for which one can approximate $R_{BCS}(T)=R_0\exp(\theta\Delta/k_BT_0)$ where $R_0=R_{BCS}(T_0)$. For $R_i=0$, Eq. (\ref{bp}) and $\partial B_p/\partial\theta=0$ can be solved exactly, giving $\theta = k_BT_0/\Delta \ll 1$, so $T_b-T_0$ and $B_b$ can be calculated in the first order in $R_i/eR_0\ll 1$:
\begin{eqnarray}
T_b-T_0=\frac{k_BT_0^2}{\Delta}\left[1+\frac{R_i}{eR_0}\right],
\label{tmm} \\
B_b=\mu_0\left[\frac{2\kappa\alpha_K k_BT_0^2}{e\Delta(\kappa+d\alpha_K)R_0}\right]^{1/2}\left[1-\frac{R_i}{2eR_0}\right],
\label{bpp}
\end{eqnarray}
where $e=2.718$. For $R_i = 5$ n$\Omega$ and $R_0=20$ n$\Omega$ at 1.3 GHz \cite{gigiJAP}, the residual resistance in Eq. (\ref{bpp}) only reduces $B_b$ by $\simeq 5\%$. The cold state 1 in Fig. 17 at $B_p<B_b$ is metastable and can be destroyed by a thermal fluctuation $\delta T \simeq T_m^{(2)}-T_m^{(1)}$, triggering thermal quench. Such  thermal bistability can result in propagation of thermal switching waves \cite{gm} or dendritic hot filaments of magnetic flux in superconducting films \cite{dendr1,dendr2}. For a clean Nb at 2 K, $\alpha_K = 5 \cdot10^3$ W/m$^2$K, $\kappa = 20$ W/mK, $\Delta/k_B = 17.5$ K, $R_s(2K) = 20$ n$\Omega$, and $d = 3$ mm, Eq. (\ref{bpp}) gives $B_b\approx 200$ mT close to $B_c$ of Nb at 0K. In this case the uniform thermal breakdown does not play a major role, and $R_s(B_p)$ is mostly controlled by nonequilibrium kinetics of quasiparticles in the Meissner state \cite{kopnin}. Thermal stability can become a problem \cite{agac} for dirtier Nb or higher-$T_c$ superconductors like Nb$_3$Sn or semi-metallic iron pnictides for which thermal conductivities are some 3 orders of magnitude lower than for Nb \cite{as,nb3sn}.

\subsection{Trapped vortex hotspots.}

We now calculate the temperature distribution around vortex hotspots in a slab.
For weak rf dissipation produced by trapped vortices, $T(x,y,z)$ is described by the linearized thermal diffusion equation:
\begin{gather}
\nabla ^{2}T=0, \label{tdif} \\
\kappa \partial _{z}T=-\tilde{P}(x,y,T),\qquad z=0 \label{bc1} \\
\kappa \partial _{z}T=-\alpha_K(T-T_{0}),\qquad z=d \label{bc2}.
\end{gather}
The surface power $\tilde P$ in the boundary condition (\ref{bc1}) includes both the rf power $R_s(T)H_p^2/2$ in the Meissner state,
and the localized heat sources $P(x,y,T)$:
\begin{equation}
\tilde{P}= P(x,y,T)+(H_p^2/2)[\partial R_s/\partial T]_{T_0}(T-T_0),
\label{tP}
\end{equation}
where $P$ can be due to a vortex hotspot or a scanning laser beam, and the second term describes the induced change
in the BCS rf heating. We consider hotspot in slabs in which $\delta T(x,y,z)$ is highly inhomogeneous
both along the surface and in the perpendicular direction, unlike the theory of scanning electron microscopy
\cite{clemh} for thin films in which $\delta T(x,y)$ is nearly uniform along $z$.

Neglecting the dynamic term $C\partial T/\partial t$ in the thermal diffusion equation implies
that dissipation is localized in a narrow layer much thinner than the thermal skin depth $\ell _{\omega }=(\kappa /C\omega )^{1/2}$,
and the rf period is much shorter than the time $t_\theta=Cd^2/\kappa$ of thermal diffusion across the film
so that temporal oscillations of $T$ are negligible, where $C$ is the specific heat. For Nb at 2 K ($\kappa =10$ W/mK,\quad $C=10^{2}$ J/m$^{3}$K), $d=2$ mm,
and $f=2$ GHz, the thermal skin depth $\ell _{\omega }=(\kappa /C\omega )^{1/2}\simeq 3\ \mu $m is much greater than $%
\lambda =40$ nm, while $t_\theta\simeq 40\ \mu$s $\gg 1/f$, justifying Eq. (2). For Nb$_{3}$Sn at 2 K ($\kappa
=10^{-2}$W/mK,\quad $C=10^{2}$J/m$^{3}$K, Ref. \onlinecite{nb3sn}) and $f=2$ GHz, $\ell _{\omega
}=(\kappa /C\omega )^{1/2}\simeq 100$ nm becomes comparable to $\lambda /2=45
$ nm.

The temperature distribution in the film can be obtained by the Fourier
transform of Eqs. (\ref{tdif})-(\ref{bc2}), as described in Appendix B:
\begin{gather}
T(\mbox{\boldmath$\rho$},\zeta)= T_0 + H_p^2R_s(T_m)d(1-\zeta+\beta\bigr)/2\kappa +
\nonumber \\
\int\frac{P_\mathbf{k}[k\beta
\cosh k(1-\zeta)+\sinh k(1-\zeta)]e^{i{\bf k}\mbox{\boldmath$\rho$}}d^2k}{\alpha_K[k\beta (1-\gamma )\cosh k+(k^2\beta^2-\gamma)\sinh k](2\pi)^2},
\label{tdis} \\
\beta = \frac{\kappa}{d\alpha_K}, \qquad \gamma=\frac{H_p^2}{2\alpha_K}\left[\frac{\partial R_s}{\partial T}\right]_{T_m},
\label{bgam}
\end{gather}
where $P_\mathbf{k} = \int e^{-i{\bf k}\mbox{\boldmath$\rho$}}P(\mbox{\boldmath$\rho$})d^2\mbox{\boldmath$\rho$}$ is the Fourier image of the localized heat source, $\mbox{\boldmath$\rho$} = {\bf r}/d$ and $\zeta = z/d$ are the dimensionless lateral and transverse coordinates, respectively, ${\bf r}=(x,y)$, and $k^2=k_x^2+k_y^2$. The second term in the right-hand side of Eq. (\ref{tdis}) is due to the uniform rf heating where
$T_m(H)$ at $z=0$ is determined by Eq. (\ref{bp}). The integral term describes a hotspot caused by the localized power source $P(x,y)$.

We now calculate the temperature disturbances $\delta T_m(\mbox{\boldmath$\rho$})=T(\mbox{\boldmath$\rho$},0)-T_m$ and $\delta T_s(\mbox{\boldmath$\rho$})=T(\mbox{\boldmath$\rho$},1)-T_s$ at the inner and outer surfaces, respectively.
Consider first $\delta T_s(\mbox{\boldmath$\rho$})$ from a heat source much smaller than the film thickness for which
$\delta T_s(\rho)$ ia axially symmetric and $P_\mathbf{k}$ can be replaced with the total power $P_0=\int dxdyP(x,y)$. Integration over the polar angle in Eq. (\ref{tdis}) yields
\begin{gather}
\delta T_s(\rho) =
\label{tempo} \\
\frac{P_0\beta}{2\pi\alpha_K}
\int_0^\infty\frac{ k^2 J_0(k\rho)dk}{k\beta(1-\gamma)\cosh k+(k^2\beta^2-\gamma)\sinh k},
\nonumber
\end{gather}
where $J_0(x)$ is the Bessel function. For $\rho=r/d\gg 1$, the integral in
Eq. (\ref{tempo}) converges at $k\ll 1$ so we can expand the denominator in $k$ and obtain (see Appendix B):
\begin{gather}
\delta T_s(r)=\frac{P_0}{2\pi
\tilde{\kappa} d}K_{0}\left(\frac{r}{\tilde{L}}\right), \quad r\gg d,
\label{tin} \\
L=\left(d\tilde{\kappa}/\tilde{\alpha}_K\right)^{1/2}, \label{L} \\
\tilde{\kappa}=\kappa\left[1+\frac{1-\gamma}{2\beta}-\frac{\gamma}{6\beta^2}\right], \label{kap} \\
\tilde{\alpha}_K=\alpha_K - \frac{H_p^2}{2}\left(1+\frac{1}{\beta}\right)\left[\frac{\partial R_s}{\partial T}\right]_{T_m},
\label{alp}
\end{gather}
where $K_{0}(x)$ is the modified Bessel function. Here $\delta T_s(r)$ decreases exponentially over the thermal length $L$ which is typically larger than the film thickness in Nb. For a slab with $d=2$ mm, $H_p=0$, $\kappa = 10$ W/mK at 2 K, we obtain $\beta = 1$ and $L=d\sqrt{\beta} = 2$ mm if $\delta T_s$ is small enough so that the superfluid He remains below $T_\lambda = 2.17$ K and $\alpha_K=5$ kW/m$^2$K. Stronger overheating, $\delta T_s > T_\lambda - T_0$, drives the liquid He above the $\lambda$-point and the Kapitza conductance drops to $\alpha_K\simeq 650$ W/m$^2$K, (Refs. \onlinecite{cryogen,kapitza}) giving $\beta = 7.7$ and $L=5.5$ mm. For a 1 $\mu$m thick Nb film and the same materials parameters, we obtain $\beta = 500$, $L\simeq 22\ \mu$m, and $\beta\simeq 4\cdot10^3$, $L\simeq 63\ \mu$m, respectively.

Equations (\ref{L})-(\ref{alp}) show that $L$ depends on the rf field amplitude because of the effect of uniform rf heating on $\tilde{\kappa}(H_p)$ and $\tilde{\alpha}_K(H_p)$. Here $L(H_p)$ increases as $H_p$ increases and diverges like $L\propto (H_b^2-H_p^2)^{-1/2}$ at the uniform breakdown field $H_b$ at which $\tilde{\alpha}_K(H)$ vanishes. The expansion of hotspots as $H_p$ increases was observed in temperature map measurements \cite{hexpan}. The rf field-induced widening of $\delta T(x,y,z)$ along the surface will be discussed below in more detail. Here we just illustrate how the expansion of hotspots can be understood from a balance of lateral thermal diffusion, rf dissipated power and the heat flux to the coolant in the region $\sim L$:
$$
d\kappa\delta T/L^2\sim [\alpha_K - (H_p^2/2)(\partial R_s/\partial T)]\delta T.
$$
Hence $L^2\sim d\kappa/[\alpha_K - (H_p^2/2)(\partial R_s/\partial T)]$ reduces to Eq. (\ref{L}) in the limit of $\beta\gg 1$ for which heat transfer is limited by the Kapitza conductance. The length $L(H_p)$ can be regarded as a thermal correlation length in the Meissner state under the rf field.

For $\beta\gg 1$, the peak value $\delta T_{s}(0)$ at the outer surface can be calculated analytically (see Appendix B):
\begin{equation}
\delta T_s(0)=\frac{P_0}{4\pi \kappa d}\ln \left( \frac{4\kappa }{d\tilde{\alpha}_K}%
\right) \label{dTs}
\end{equation}
Eq. (\ref{dTs}) overlaps with Eq. (\ref{tin}) at $r\sim d$ in which $K(x)\simeq \ln (1/x)$ at $x\ll 1$. Notice that $\delta T_s(0)$ depends only weakly on the
Kapitza thermal resistance.

\subsection{Reconstruction of heat sources from temperature maps.}

The results presented above enable one to reconstruct the distribution of local power sources $P(\mathbf{r})$ from the temperature maps of $\delta T(\mathbf{r})$, using Eq. (\ref{tdis}) which links the Fourier components $P_{\mathbf{k}}$ and $\delta T_{\mathbf{k}}$:
\begin{gather}
P_{\mathbf{k}}=
\alpha_K\!\left[(1-\gamma)\cosh k+\left( k\beta -\frac{\gamma}{k\beta }\right) \sinh k\right]\!
\delta T_{\mathbf{k}}
\label{fur}
\end{gather}
The formally exact Eq. (\ref{fur}) cannot be directly applied to reconstruct $P(\mathbf{r})$ using the fast Fourier transform of the
measured $\delta T(\mathbf{r})$, because the hyperbolic functions in Eq. (\ref{fur}) greatly amplify the contribution of
short wavelength harmonics of inevitable noise in the measured signal. This problem is resolved using the standard methods of reducing the signal-to-noise ratio in which the measured $\delta T(x,y)$ distribution should be first coarse-grained to remove all Fourier components of fictitious
temperature fluctuations with the periods shorter than the spatial resolution of the thermal map measurements \cite{imag1,imag2}. The spatial resolution of our thermal maps $\ell_m\simeq 1$ cm does not allow probing the length scales $\sim d\simeq 2-3$ mm, so Eq. (\ref{fur}) should be expanded in small $k$ up to terms $\sim k^2$. Restoring the normal units yields
\begin{equation}
\overline{P}_{\mathbf{k}}=\tilde{\alpha}_K(1+L^2k^2)\overline{\delta T}_{\mathbf{k}},
\label{fur1}
\end{equation}
where $L$ and $\tilde{\alpha}_K$ are defined by Eqs. (\ref{L}) and (\ref{alp}).
The wavevector $\mathbf{k}$ is restricted by the condition $k<k_0\simeq \ell_m^{-1}$ and the over line implies spatial averaging of the measured
$\delta T(x,y)$ which eliminates all harmonics with $k>k_0$. In the coordinate space Eq. (\ref{fur1}) takes the form
\begin{equation}
\overline{P}(\mathbf{r})=\tilde{\alpha}_K[\overline{\delta T}(\mathbf{r})-L^2\nabla ^2\overline{\delta T}(\mathbf{r})].
\label{reconstr}
\end{equation}
This equation can be used to reconstruct the distribution of power sources $\overline{P}_v(x,y)$ from the smoothed thermal maps
$\overline{\delta T}(x,y)$. For instance, the observed $\overline{\delta T}(x,y)$ which can be approximated by Eq. (\ref{tin}) up to $r\simeq \ell_m$,
suggests a small heat source of size $r_0 \ll L$ for which the total power $P_0$ but not $r_0$ can be measured.

Now we estimate the number of trapped vortices which can produce the observed peaks in $\delta T_s \simeq 0.2-0.5 $ K shown in Fig. 3. For intermediate frequencies $\omega > \omega_\ell$ relevant to our experiment, Eq. (\ref{P2}) yields $P\simeq 0.07\ \mu$W per vortex at $B_p=74$ mT and 2 GHz. To see how many vortices can produce $\delta T \simeq 0.2-0.5$ K in the Nb plate of thickness 2 mm, we consider two limits of a localized vortex bundle with $r_0\ll L$ and a distributed bundle with $r_0\gtrsim L$. For a localized bundle, we use Eq. (\ref{dTs}) with $\kappa =10$ W/mK, $d=2$ mm, $\alpha_K=650$ W/m$^2$K. Then we obtain that $P_0=4\pi \kappa d\delta T_s(0)/\ln (4\kappa/d\alpha_K)\simeq 15-37$ mW for $\delta T_s(0) = 0.2-0.5$ K. This requires $N=P_0/P\sim (2-5)\cdot10^5$ vortices. If they are spaced by distances $\sim\lambda = 40$ nm, the size of the vortex bundle $\sim \lambda N^{1/2}\sim 20-30\ \mu$m is much smaller than $d$.

To estimate the density $n_v$ of trapped vortices which can produce $\delta T_s \simeq 0.2-0.5$ K in a distributed bundle, we use the uniform heat balance condition, $\delta T_s\kappa\alpha_K/(\kappa +d\alpha_K)\simeq n_v P$. Hence, $n_v\sim \alpha_K\kappa\delta T_s/P(\kappa+d\alpha_K)\sim (2-5)\cdot10^9$ m$^{-2}$ corresponds to the mean distance between vortices $l_v=n_v^{-1/2}\sim 14-20\ \mu$m, and the effective magnetic induction $B_v=\phi_0n_v \simeq 4-10\ \mu$T, smaller than 25-60 $\mu$T of the unscreened Earth magnetic field. Such local variations of the vortex density may result from mesoscopic fluctuations of random pinning forces.

\subsection{Temperature distribution at the inner surface.}

The distribution of $\delta T_m(r)$ at the inner
surface follows from Eq. (\ref{tdis}) at $\zeta=0$:
\begin{gather}
\delta T_m(\rho)=\frac{1}{2\pi\alpha_K}\!\int_{0}^{\infty }\!\!\frac{(k\beta+\tanh k)P_\mathbf{k}J_{0}(k\rho)kdk}{k\beta (1-\gamma)+(k^2\beta^2-\gamma)\tanh k}
\label{inT}
\end{gather}
For a point heat source, this integral diverges at $\rho=0$, so we consider a more realistic Gaussian distribution $P(r)=(P_0/2\pi r_{0}^{2})\exp (-r^{2}/2r_{0}^{2})$
(in real units) and the Fourier transform $P(p)=P_0\exp (-p^{2}r_{0}^{2}/2)$ where $k=pd$, and $P_0$ and $r_{0}<d$ are the total power and the radius of the source, respectively. Such $P(r)$ can model both a trapped vortex bundle or a laser beam.

For $r<d$, the main contribution to the integral in Eq. (\ref{inT}) comes from the region of $k>1$ where $\tanh k\approx 1$.
Then $\delta T_m(r)$ can be calculated analytically if $\beta\gg 1$:
\begin{eqnarray}
\delta T_m(r)=\frac{P_0}{2\pi \kappa }\int_{0}^{\infty }\exp
(-p^{2}r_{0}^{2}/2)J_{0}(pr)dp =\nonumber \\
\frac{P_0}{2\sqrt{2\pi }\kappa r_{0}}\exp \left( -\frac{r^{2}}{
4r_{0}^{2}}\right) I_{0}\left( \frac{r^{2}}{4r_{0}^{2}}\right), \qquad r<d,
\label{tm}
\end{eqnarray}
where $I_{0}(x)$ is the modified Bessel function. The peak value $\delta T_m(0)$ is given by
\begin{equation}
\delta T_m(0)=\frac{P_0}{2\sqrt{2\pi }\kappa r_{0}}
\label{dTim}
\end{equation}
For $r>2r_{0}$, the asymptotic expansion of $I_{0}(x)=\exp (x)/\sqrt{2\pi x}$ in Eq. (\ref{dTim}) yields the
temperature distribution from a point source in a semi-infinite media:
\begin{equation}
\delta T_m(r)=\frac{P_0}{2\pi \kappa r},\qquad r>2r_{0}
\label{dTip}
\end{equation}
The maximum temperature gradient
\begin{equation}
|\nabla\delta T|_{max}=0.071P_0/\kappa r_0^2
\label{tgrad}
\end{equation}
occurs at $r=1.194r_0$. Using Eqs. (\ref{dTs}) and (\ref{tgrad}), we obtain the relation between the peak values of $\delta T$ on
the inner and outer surfaces:
\begin{gather}
\delta T_m(0) = \frac{\sqrt{2\pi}d \delta T_s(0)}{r_0\ln(4\kappa/d\tilde{\alpha}_K)}
\label{ms} \\
|\nabla\delta T|_c\simeq 0.36T_m(0)/r_0
\label{msg}
\end{gather}
For $r_0=0.25$ mm, $d=2$ mm, $\alpha_K= 650$ W/m$^3$K, $\kappa=10$ W/mK, and $\delta T_s(0)=0.3$ K,
Eqs. (\ref{ms}) and (\ref{msg}) yield $\delta T_m(0)\simeq 1.8$ K and $|\nabla\delta T|_c\simeq 2.6$ K/mm.
Equations (\ref{dTip}) and (\ref{tgrad}) can be used to evaluate the temperature gradient produced by
scanning laser beam for which $P_0=\alpha_\omega W$, $W$ is the laser power, and $\alpha_\omega$ is the absorption coefficient.
Substituting Eq. (\ref{tgrad}) into Eq. (\ref{nT}) then gives the minimum beam power $W_c$ to move trapped vortices
\begin{equation}
W_c\sim 232\kappa\mu_0J_c r_0^2\lambda_0^2T_c^2/\alpha_\omega\phi_0T.
\label{cw}
\end{equation}
The critical power $W_c$ can be reduced by focusing the laser beam to diminish $r_0$ in Eq. (\ref{cw}).

\begin{figure}
\centerline{\psfig{file=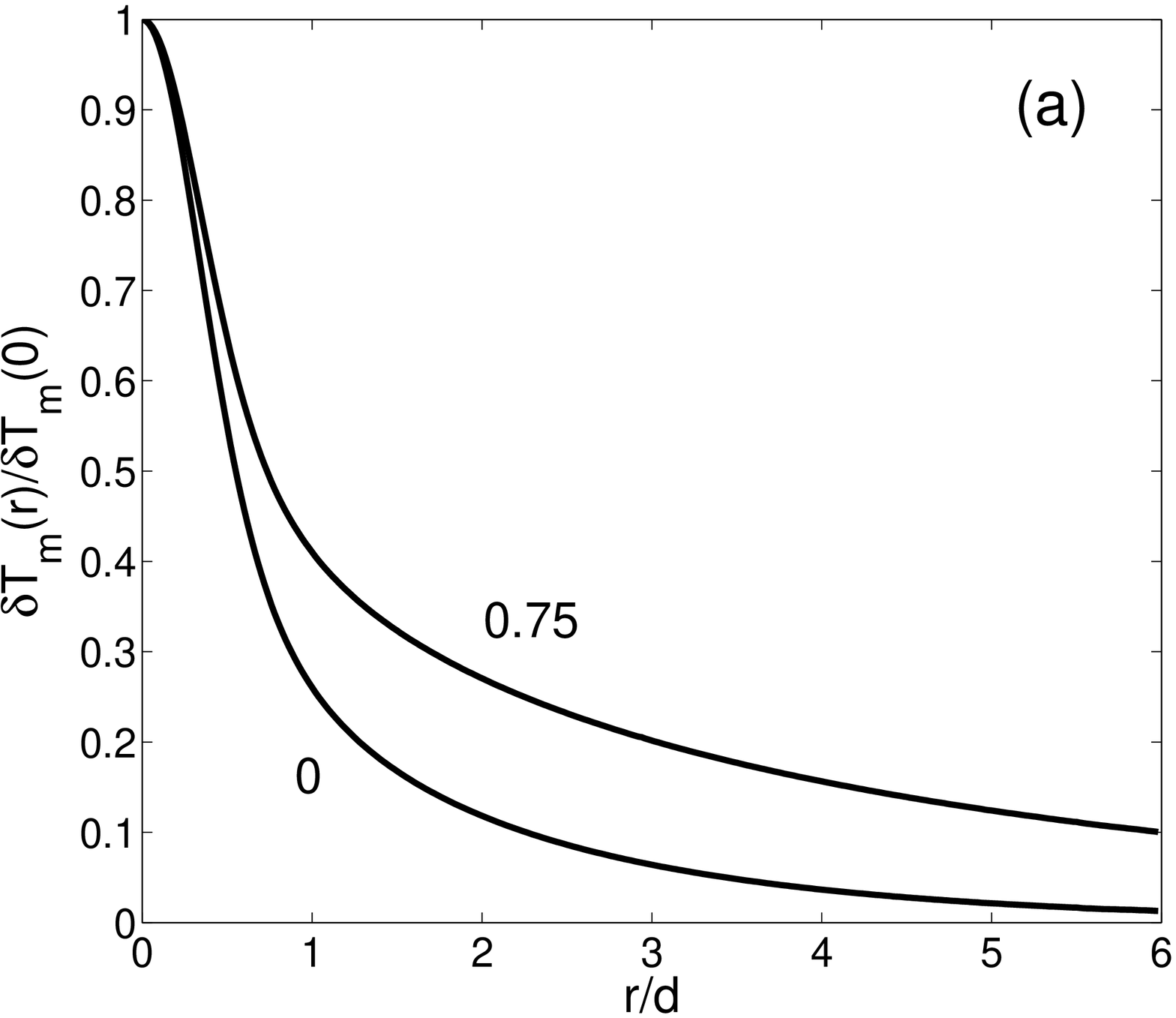,width=6cm}}
\vspace{3mm}
\centerline{\psfig{file=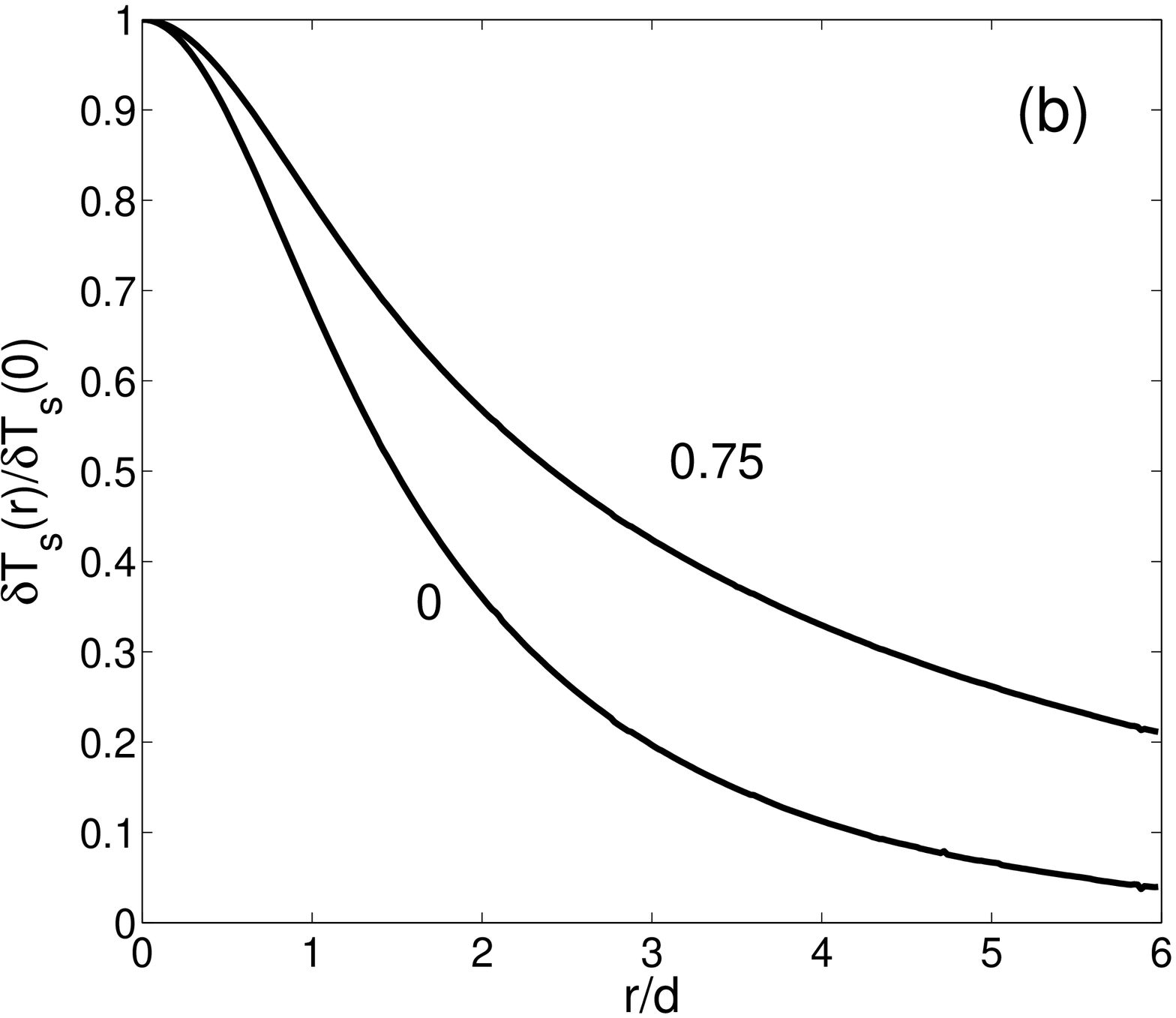,width=6cm}}
\caption{Lateral distributions of $\delta T_m(r)$ and $\delta T_s(r)$ normalized to their respective peak values calculated from
Eq. (\ref{inT}) and (\ref{tempo}) for the Gaussian power source $P_k=P_0\exp(-a^2k^2)$, $a=0.2d$, and $\beta = 5$. The lower and the upper curves
in both (a) and (b) correspond to $\gamma=0$ and $\gamma=0.75$, respectively. }
\label{fig18}
\end{figure}

Shown in Fig. 18 are the features of $\delta T_s(r)$ and $\delta T_m(r)$ calculated numerically from
Eqs. (\ref{tempo}) and (\ref{inT}) for a Gaussian heat source with $r_0\ll d$ and different values of the uniform heating parameter $\gamma(H_p)$. One can clearly
see the widening of the hotspot as $H_p$ increases, as was mentioned above. Another feature is the distinctly different behaviors of
$\delta T_m(r)$ and $\delta T_s(r)$ at small distances $r<d$ from the power source: $\delta T_m(r)$ has a sharp peak at $r<d$ described by Eqs. (\ref{tm})-(\ref{dTip}),
but $\delta T_s(r)$ is rounded by thermal diffusion across a thick sample. For large distances $r\gg d$, the temperature across the film becomes nearly uniform in the limit of strong thermal diffusion, $\beta = \kappa/d\alpha_K\gg 1$, and $\delta T_m(\rho)$  becomes equal to $\delta T_s(\rho)$. However,
near the heat source, the local overheating at the inner surface $\delta T_m(r)$ can be much higher than its "image" $\delta T_s(r)$ at the outer surface.

\subsection{Scanning laser microscopy of hotspots.}

Scanning laser can be used to probe local inhomogeneity of the surface resistance $R_s(\mathbf{r})$. If the laser beam locally increases the temperature by $\delta T_m(x,y)<T_0^2/T_c$, the surface resistance changes by $\delta R(\mathbf{r}) = [\partial R_s(\mathbf{r}, T)/\partial T]_{T_0}\delta T_m(\mathbf{r-r_0})$, where $\mathbf{r_0}$ is the position of the beam at the surface. As a result, the quality factor at low rf fields decreases by
\begin{equation}
\delta Q(\mathbf{r_0}) = \frac{Q}{A\overline{R}_s}\int \left[\frac{\partial R_s(\mathbf{r}, T)}{\partial T}\right]_{T_0}\delta T_m(\mathbf{r-r_0})d^2{\mathbf r},
\label{dQ}
\end{equation}
where $A$ is the total surface area, $\overline{R}_s$ is the averaged surface resistance, and $\delta T_m(\mathbf{r})$ given by Eqs. (\ref{inT})-(\ref{tm}) decreases as $1/r$ for $r<d$ and as $\exp(-r/L)$ for $r>L$. If $R_s(\mathbf{r})$ varies slowly over the thermal length $L$,
the derivative $[\partial R_s(\mathbf{r}, T)/\partial T]_{T_0}$ can be taken out of the integral in Eq. (\ref{dQ}) which then reduces to the zero-$k$
Fourier component $\delta T_m(k=0)= (1/\alpha_K+d/\kappa)P_0$, as follows from Eq. (\ref{inT}). Thus,
Eq. (\ref{dQ}) becomes:
\begin{equation}
\delta Q(\mathbf{r_0}) = \frac{QP_0}{A\overline{R}_s}\left(\frac{1}{\alpha_K}+\frac{d}{\kappa}\right)\left[\frac{\partial R_s(\mathbf{r_0}, T)}{\partial T}\right]_{T_0}
\label{dQs}
\end{equation}
Now we estimate $\delta Q/Q$ due to trapped vortex bundles which locally increase $R_i(\mathbf{r_0})$. Evaluating $\partial R_s/\partial T\simeq 2TR_i(\mathbf{r_0})/T_c^2$ from Eq. (\ref{ri}) with $B_c(T)=B_c(0)(1-T^2/T_c^2)$, yields
\begin{equation}
\frac{\delta Q(\mathbf{r_0})}{Q}\simeq \left(\frac{2T_0P_0}{AT_c^2}\right)\left(\frac{1}{\alpha_K}+\frac{d}{\kappa}\right)\left[\frac{R_i(\mathbf{r_0})}{\overline{R}_s}\right]
\label{estQ}
\end{equation}
For the parameters of our experiment, $P_0\simeq 1$W, $A\simeq 10^{-2}$ m$^2$, $\alpha_K\simeq 650$ W/m$^2$K, $d=2$ mm, $\kappa=10$ W/mK, $T_0=2$ K and $T_c=9.2$ K, Eq. (\ref{estQ}) shows  that, if vortex hotspots locally increase the residual resistance to $R_i(\mathbf{r_0})\sim 10^2\overline{R}_s$, the global quality factor would change by $\delta Q\sim Q$. Thus, laser scanning of comparatively weak vortex hotspots with $R_i(\mathbf{r_0})\sim (1-10)\overline{R}_s$ can result in detectable (a few $\%$) change in the global $Q$.

\section{Effect of hotspots on surface resistance.}

Trapped vortices give rise to two different contributions to $R_s$. The first
one is given by Eqs. (\ref{Ri}) and (\ref{ri}), and another one comes from the increase of $R_s(T)$ caused by local
overheating in hotspots. Generally, the calculation of non-isothermal $R_s$ with randomly-distributed hotspots
requires solving a highly nonlinear partial differential equation for $T(x,y)$ with inhomogeneous parameters. We consider here
two simpler limits of sparse weak vortex hotspots, and overlapping hotspots spaced by distances much smaller than the thermal
length $L$.

\subsection{Sparse hotspots}

The rf power $P_t$ generated by a trapped vortex bundle consists of two contributions:
\begin{gather}
P_t=P_0 + \frac{H_p^2}{2}\left[\frac{\partial R_s}{\partial T}\right]_{T_m}\int\delta\tilde{T}(x,y,0)dxdy = \nonumber \\
=P_0 + \frac{H_p^2(1+\beta)P_0}{2\alpha_K[\beta - (\beta+1)\gamma(H_p)]}\left[\frac{\partial R_s}{\partial T}\right]_{T_m}
\label{pt}
\end{gather}
Here $P_0$ is the rf power directly dissipated by a vortex bundle, and the integral term accounts for the induced
increase of the surface resistance in a surrounding warmed-up area. The integral is the Fourier
component of $\delta\tilde{T}(\mathbf{k})$ at ${\mathbf k}=0$ given by the last term in Eq. (\ref{tdis}) at $\zeta=0$.

For noninteracting hotspots spaced by distances $>L$, the global dissipated rf power $\sum_n P_t({\bf r_n})$ is just a sum of
contributions of hotspots located at ${\bf r_n}$. If we associate the global residual resistance $\overline{R}_i$ only with vortex
hotspots, then $H_p^2{\overline R}_i/2 =\sum_n P_t({\bf r_n})$. Using Eq. (\ref{pt}) and Eq. (\ref{bgam}) for $\beta$ and $\gamma(H_p)$, we obtain
\begin{equation}
\overline{R}_i=\frac{R_i}{1 - (H_p^2/2)(1/\alpha_K+d/\kappa)[\partial R_s/\partial T]_{T_m}}.
\label{risp}
\end{equation}
The term $\propto H_p^2$ in the denominator of Eq. (\ref{risp}) makes the residual resistance dependent on the
rf field amplitude due to the field-induced expansion of hotspots, as
was discussed above. Moreover, Eq. (\ref{risp}) predicts that $\overline{R}_i(H_p) \propto R_i/(1-H_p^2/H_b^2)$
would diverge as $H_p$ approaches the uniform thermal breakdown field $H_b$. In this case, the assumption of non-overlapping
hotspots fails, and the calculation of $\overline{R}_i$ should take thermal interaction of hotspots into account.

Equation (\ref{risp}) shows that the rf heating not only gives rise to the field dependent $\overline{R}_i(H_p)$, it also makes the residual resistance interconnected with the BCS surface resistance contributing to $\partial R_s/\partial T = (\Delta/k_BT^2)R_{BCS}+2TR_i/T_c^2$ in the denominator of Eq. (\ref{risp}). Here
$\partial R_i/\partial T \simeq 2TR_i/T_c^2$ was evaluated using Eq. (\ref{ri}) with $H_c(T)=H_c(0)(1-T^2/T_c^2)$. Therefore, separation of $R_s$ into the BCS and the residual contribution is well-defined only at weak fields for which heating is negligible. The field dependence of $R_i(H_p)$ due to sparse vortex bundles can significantly reduce the quality factors of the Nb resonator cavities at intermediate and high rf fields \cite{hexpan}.

\subsection{Overlapping hotspots}

For overlapping hotspots, we define the local residual resistance $R_i({\bf r})=\overline{R}_i+\delta R({\bf r})$ where
$\overline{R}_i$ is the mean value resulting from contributions of all vortex hotspots, and the random variations $\delta R({\bf r})$ are
due to mesoscopic fluctuations of pinning forces and the cooling pre-history of the sample,
as was described above. Here the random variable $\delta R({\bf r})\propto \delta n_v({\bf r})$ has zero mean $\langle\delta R\rangle=0$ and
is proportional to the local fluctuation of the vortex density $\delta n_v({\bf r})$. Spatial fluctuations of $\delta R$ are characterized
by the correlation function $F(|{\bf r} - {\bf r'}|) = \langle \delta R({\bf r})\delta R({\bf r'})\rangle$:
\begin{equation}
F(|{\bf r} - {\bf r'}|) =
\langle \delta n_v({\bf r})\delta n_v({\bf r'})\rangle {\overline R}_i^2/\overline {n}_v^2,
\label{F}
\end{equation}
where $\langle ... \rangle$ means statistical averaging, $\overline{n}_v$ is the mean density of trapped vortices, and $F(|{\bf r} - {\bf r'}|)$ is  proportional to the correlation function of density fluctuations of randomly distributed vortex bundles. We assume that fluctuations $\delta R({\bf r})$ are isotropic along the surface so that $F(|{\bf r} - {\bf r'}|)$ depends only on ${\bf r}-{\bf r'}$. The following calculation of the global surface resistance $\overline{R}_s$ is valid for any form of $F({\bf r}-{\bf r'})$, but specific formulas will be obtained for the conventional Gaussian function
\begin{equation}
F(|{\bf r} - {\bf r'}|)=\langle\delta R^2\rangle\exp\left[-\frac{|{\bf r} - {\bf r'}|^2}{r_c^2} \right],
\label{gauss}
\end{equation}
where $\langle\delta R^2\rangle^{1/2}$ and the correlation radius $r_c \ll L$ quantify characteristic magnitudes and spatial scales of local fluctuations of $\delta R$. For sparse hotspots, $r_c$ is of the order of the mean spacing between vortex bundles. The global surface resistance $\overline{R}_s$ is then,
\begin{equation}
\overline{R}_s = R_{BCS}(T_m) +\overline{R}_i + \frac{\langle\delta T^2\rangle}{2}\left[\frac{\partial^2R_s}{\partial T^2}\right]_{T_m}.
\label{rss}
\end{equation}
Here $\partial^2 R_s/\partial T^2\simeq (\Delta/k_BT)^2R_{BCS}(T) + \partial^2 \overline{R}_i/\partial T^2$, and $\delta T({\bf r})$ are random temperature fluctuations around the mean $T_m$ defined self-consistently by the equation:
\begin{equation}
H_p^2\overline{R}_s(T_m, H_p) = 2\kappa\alpha_K(T_m-T_0)/(d\alpha_K + \kappa).
\label{hbr}
\end{equation}
To obtain $\langle \delta T^2\rangle$ in Eq. (\ref{rss}) we use Eq. (\ref{tdis}) in which $\zeta = 0$, and $P({\bf k})=H_p^2\delta R({\bf k})/2$ is the
fluctuation dissipation, and $\langle P({\bf k})P({\bf k'})\rangle = (H_p^4/4)F({\bf k})\delta ({\bf k}+{\bf k'})$. Hence,
\begin{equation}
\langle\delta T^2\rangle=\left[\frac{H_p^2}{4\pi\alpha_K}\right]^2\!\int\frac{F({\bf k})(k\beta
 +\tanh k)^2d^2{\bf k}}{[k\beta (1-\gamma )+(k^2\beta^2-\gamma)\tanh k]^2},
\label{dt2}
\end{equation}
where $F({\bf k})\delta ({\bf k}+{\bf k'})$ is the Fourier image of $F(|\mbox{\boldmath$\rho$} - \mbox{\boldmath$\rho'$}|)$. We consider here the limit of
$\beta\gg 1$ for which $\langle\delta T^2\rangle$ can be evaluated analytically, using $F(k)=\pi \varrho^2\langle\delta R^2\rangle\exp(-\varrho^2k^2/4)$ with $\varrho = r_c/d \ll 1$ for the Gaussian correlation function. In this case the main contribution to the integral in Eq. (\ref{dt2}) comes from the region of $k\ll 1$ where $\tanh k=k$ and the upper limit can be extended to $\infty$, as shown in Appendix B. This gives $\langle \delta T^2\rangle$ which essentially depends on $T_0$, $H_p$ and $\omega$:
\begin{equation}
\langle\delta T^2\rangle=\frac{H_p^4\langle\delta R^2\rangle r_c^2}{16d\kappa\alpha_K[1-\gamma(H_p,\omega)]}, \qquad \beta\gg 1.
\label{dtb}
\end{equation}
At $T\ll T_c$, the variance  $\langle\delta R^2\rangle$ is temperature independent, but $\kappa\propto T^3$ and $\alpha_K\propto T^n$ with $n=3-5$ are mostly determined by the phonon heat transport \cite{cryogen}. Thus, $\langle\delta T^2\rangle$ increases as $T$ decreases. The dependence of $\langle \delta T^2\rangle$ on $H_p$ varies from $\langle\delta T^2\rangle \propto H_p^4$ at $H_p\ll H_b$ to a much stronger increase at $H_p\simeq H_b$ as the factor $1-\gamma(H_p)$ in the denominator of Eq. (\ref{dt2}) diminishes. The latter reflects the divergence of temperature fluctuations as $H_p$ approaches the field of uniform thermal instability at which $1=\gamma(H_p)$. Equations (\ref{rss}) and (\ref{dtb}) yield
\begin{gather}
\overline{R}_s = R_s(T_m) + \frac{H_p^4\langle\delta R^2\rangle r_c^2}{32d\kappa\alpha_K[1-\overline{\gamma}(H_p)]}\left[\frac{\partial^2R_s}{\partial T^2}\right]_{T_m},
\label{rsg} \\
R_s=R_{BCS}+\overline{R}_i, \qquad \overline{\gamma}=\frac{H_p^2}{2\alpha_K}\left[\frac{\partial R_s}{\partial T}\right]_{T_m}.
\label{rsgam}
\end{gather}
Here both $R_{BCS}(T_m)$ and $\overline{R}_i(T_m)$ depend on the mean surface temperature $T_m$ defined self-consistently by Eqs. (\ref{hbr}), (\ref{rsg}), and (\ref{rsgam}).

\begin{figure}
\centerline{\psfig{file=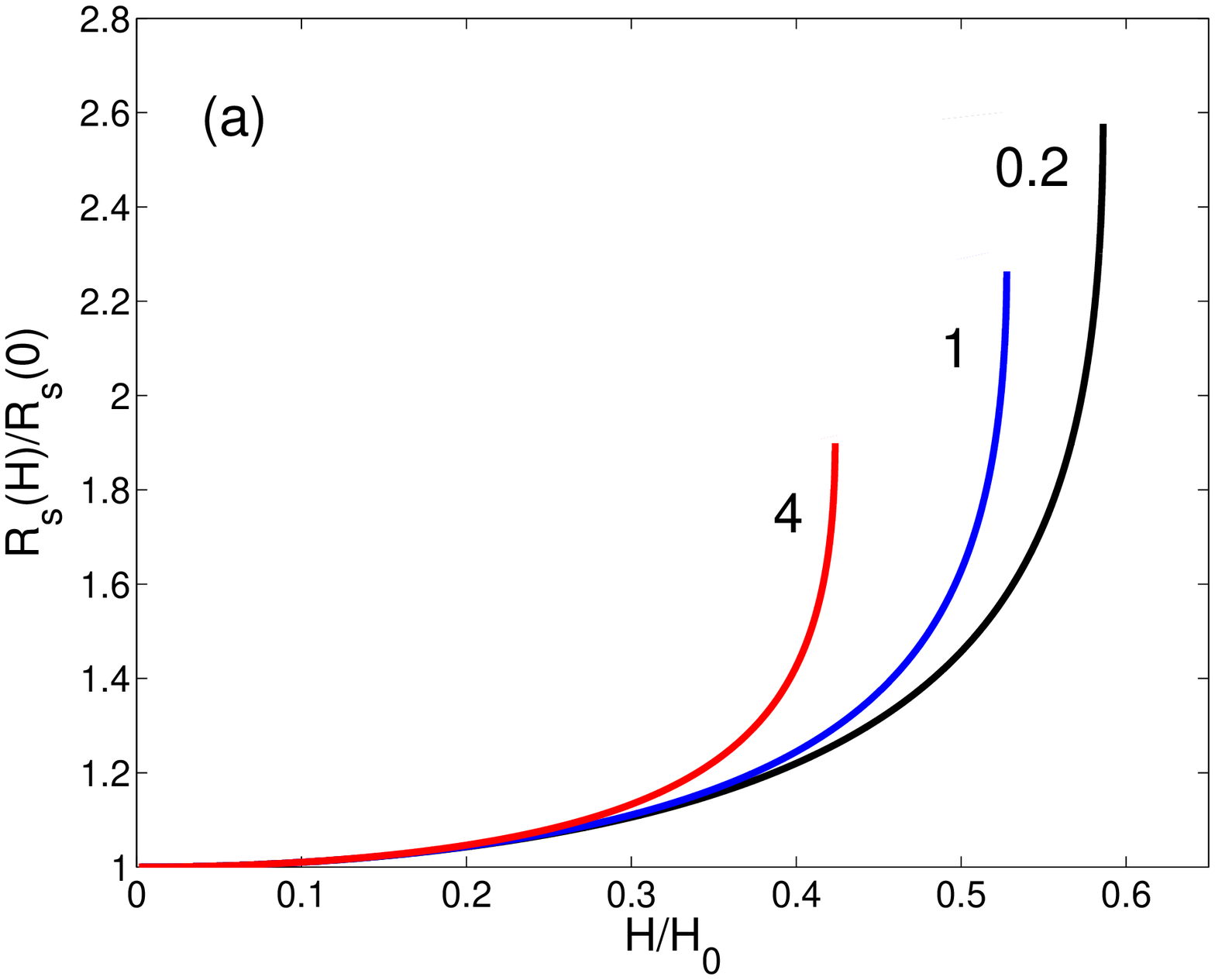,width=6.5cm}}
\centerline{\psfig{file=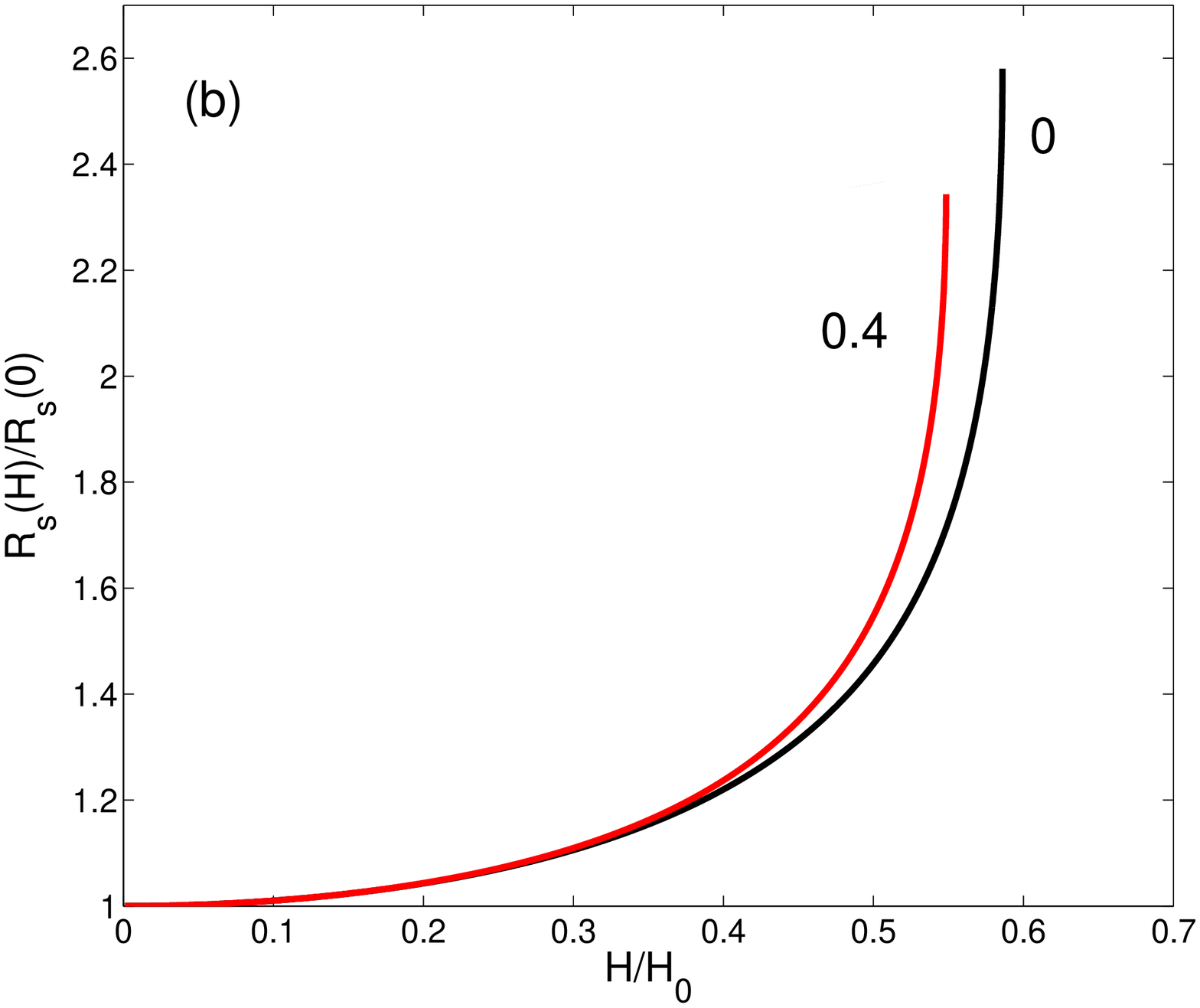,width=6.1cm}}
\caption{(a) Surface resistance calculated from Eqs. (\ref{hbr})-(\ref{dt2}) disregarding spatial variations of the vortex density for
$R_i/R_0=0.2, \  1, \ 4$. The endpoints of all $R_s(H)$ curves correspond to the rf field amplitudes at which thermal instability occurs and the slopes of $\partial R_s/\partial H$ diverge. (b) Effect of spatial variations of the hotspot distribution on the field dependence of $\overline{R}_s$ described by Eqs. (\ref{ttt}) and (\ref{rso}) for $R_i = 0.25 R_0$ and two different parameters $s=0$ and $0.4$. The field is normalized to
$H_0=(2\alpha_K k_BT_0^2/\Delta R_0)^{1/2}$.}
\label{fig19}
\end{figure}

If $R_{BCS}(T)\gg (T/T_c)^3 R_i$, Eqs. (\ref{hbr}) and (\ref{rsg}) can be written in the following dimensionless form
\begin{gather}
\theta = h^2\left[e^\theta + r+\frac{sh^4e^\theta}{1-h^2 e^\theta}\right],
\label{ttt} \\
R_s = R_0\theta/h^2, \qquad s=\langle\delta R^2\rangle r_c^2/32 R_0^2L^2,
\label{rso}
\end{gather}
where $h=H_p/H_0$, $H_0=2\alpha_Kk_BT_0^2/\Delta R_0$, $r=R_i/R_0$, and $\theta=(T_m-T_0)\Delta/k_BT_0^2$. Equation (\ref{ttt}) is a cubic equation for $h^2(\theta)$ from which the breakdown field is determined by the condition $\partial h/\partial\theta = 0$.

Shown in Fig. 19 are the dependencies of $\overline{R}_s$ on the rf field amplitude calculated from Eqs. (\ref{ttt}) and (\ref{rso}).
If the effect of local fluctuations $\delta R({\mathbf r})$ is negligible $(s\rightarrow 0)$,
Eq. (\ref{ttt}) yields the explicit relation $h^2=\theta/(r+e^\theta)$ from which $R_s(h)$ can be calculated for different values of
$r=R_i/R_0$ (Fig. 19a). Here $R_s$ increases with the rf field amplitude, while the increase of $R_i$ reduces the thermal breakdown field
and flattens the $R_s(H)$ curves. For instance, the ratio of $R_s(H_b)/R_s(0)$ decreases from
$e=2.718$ at $r=0$ to $\approx 1.9$ at $r=4$. Fig. 19b shows the effect of spatial correlations of fluctuations of vortex density on
$\overline{R}_s(H_p)$ as the parameter $s$ increases from $0$ to $0.4$.

\section{Discussion}

The results of this work show that the hotspots in the Nb plate observed in our temperature map measurements
are due to trapped vortex bundles which can be moved and broken in pieces by the scanning laser beam. The scanning
laser can be used both to reveal the 2D map of the vortex hotspots in the LSM mode and as a
"thermal broom" pushing trapped vortices out of the sample to increase the quality factor.
Moving vortices out of resonator cavities would require displacing them over long distances $(\sim 1-10)$ cm, all
the way to the cavity orifices, or annihilating trapped vortex loops at the surface by strong local overheating and temperature
gradient produced by the laser beam. The scanning laser may remove trapped vortices
more effectively in thin film structures where it can push much shorter $\sim 10^{-2}-1\ \mu$m
perpendicular vortices toward the film edges.

Vortices can significantly contribute to the residual surface resistance: as was shown above, even a very low
density of trapped vortices corresponding to $\sim 1 \%$ of the Earth magnetic field can result in $R_i\sim 1$ n$\Omega$ at 2 K and 2 GHz in magnetically
screened high purity Nb. This conclusion is consistent with the fact that some of the hotspots observed by thermal maps are indeed due to trapped vortices.
Our calculations of the power dissipated by trapped vortex segments show that $P(\omega, l_i)$ is a complicated function of
the rf frequency $\omega$ and the mean free path $l_i$, suggesting that reducing vortex dissipation
can require making the either material cleaner or dirtier, depending on the particular frequency range. At the same time, increasing
the density of pinning centers reduces $R_i$ at low fields, $H_p\ll H_c$.

Trapped vortices, along with intrinsic mechanisms of the rf pairbreaking and  nonequilibrium kinetics of quasiparticles \cite{kopnin}, can
contribute to the nonlinearity of the electromagnetic response. One of the mechanisms is due to local overheating in vortex hotspots, reducing the
breakdown field of low-dissipative Meissner state and igniting thermal quench propagation along the sample surface. Vortex contribution
can become even more essential in the rf resonator structures using superconductors with $T_c$ and $H_c$ higher than Nb, for example, Nb$_3$Sn, MgB$_2$, or iron pnictides. Because such materials have much lower thermal conductivities \cite{nb3sn,as} and smaller $H_{c1}$, reducing local overheating of trapped vortex bundles may require surface multilayer nanostructuring of Nb cavities to increase $H_{c1}$ without impeding heat transfer \cite{agac,ml}.

At higher fields, $H_p>H_{pin}$ the Meissner currents can tear vortex segments off defects at the surface so that the tips of the vortices shown in Figs. 1 and 14 get depinned, while the rest of the threading vortices remain pinned. For a nano-precipitate spaced by $\ell<\lambda$ from the surface, the depinning field $H_{pin}$ estimated from Eq. (\ref{hpin}) is much smaller than $H_c$. As a result, trapped vortex bundles can cause microwave nonlinearities at the fields $H_p\ll H_c$ much smaller than the rf fields $H_p\simeq H_c$ which have been reached in the Nb resonator cavities \cite{cav,cav1,agac}. The "unzipping" of the trapped vortex tips from the pins can result in the generation of higher harmonics \cite{3harm}, and jump-wise instabilities \cite{gurci} of rapidly moving vortex segments at the field amplitudes of the order of the superheating field $H_{sh}\simeq H_c$, Ref.  \onlinecite{sh}. Thus, even the most effective core pinning can hardly reduce $R_i$ at $H_p\simeq H_c$: as follows from Eq. (\ref{hpin}), reaching $H_{pin}\simeq H_c$ would require such high density of the optimal pinning defects of radius $\simeq \xi$ spaced by $\ell\simeq\xi$ that they would block the Meissner screening currents and reduce $\Delta$ due to the proximity effect \cite{book}. The microstructural surface analysis of the high-performance Nb resonator cavities shows a much lower density $(\ell\gg\xi)$ of pinning defects \cite{cav1}.

\begin{acknowledgments}
Funding for this work was provided by American Recovery and Reinvestment Act (ARRA) through the US Department of Energy.
\end{acknowledgments}

\appendix

\section{Dissipated power}
The mean power $P = \langle\int_0^\ell \dot{u}(z,t)F(z,t)dz\rangle_\omega$ dissipated by a vortex segment can be written in the form
$P=-(\omega/2)\mbox{Im}\int_0^\ell u(z,\omega)F(z,-\omega)dz$, where $F(-\omega,z) = Fe^{-z/\lambda}$ is
the Lorentz force, and $F = H_p\phi_0/\lambda$. Using here $u(z,\omega)$ from Eq. (\ref{u}), we obtain:
\begin{gather}
P=-\frac{F\omega}{2}\mbox{
Im} \sum_{n}A_n\int_0^{\ell }\cos (k_n z)\exp (-x/\lambda )dz
\nonumber \\
=\frac{1}{4}\sum_{n}\frac{F^2\omega ^{2}\eta
I_n^2\ell}{\omega ^2\eta ^2+k_n^4\varepsilon ^{2}(k_{n})},
\label{A1} \\
I_{n}=\frac{2}{\ell}\int_0^{\ell }\cos (k_n z)e^{-x/\lambda}dz =
\nonumber \\
\frac{\pi (2n+1)(-1)^{n}e^{-a}+2a}{a^{2}+\pi ^{2}(n+1/2)^{2}},
\label{A2}
\end{gather}
where $a=\ell/\lambda$. For a long segment, $\ell >\lambda $, the main contribution to the sum in Eq. (\ref{A1}) comes from
$k_{n}\lambda <1$ for which we can neglect the $k$-dependence of $\varepsilon (k)$ and set $e^{-a}\rightarrow 0$ in Eq. (\ref{A2}). Then
\begin{eqnarray}
P=
\sum_{n=0}^{\infty }\frac{P_1}{[\nu^{2}+\pi ^{4}(n+1/2)^{4}][a^{2}+\pi
^{2}(n+1/2)^{2}]^{2}},
\label{A3}
\end{eqnarray}
where $P_1=F^{2}\ell ^{7}\omega ^{2}\eta/\varepsilon ^{2}\lambda ^{2}$ and $\nu=\omega \eta \ell ^{2}/\varepsilon $.
The sum in Eq. (\ref{A3}) is then
\begin{gather}
S=\frac{\mbox{Re}}{4a^{3}\nu^{3/2}(a^{4}+\nu^{2})^{2}}\biggl[\nu^{7/2} -2a^{7}\sqrt{i}\tan\sqrt{i\nu}- \nonumber \\
4a^{5}\nu\sqrt{-i}\tan\sqrt{i\nu}+2a^{3}\nu^{2}\sqrt{i}\tan\sqrt{i\nu
}+5a^{4}\nu^{3/2}\biggr].
\label{A4}
\end{gather}
Using
\begin{equation}
\mbox{Re}\sqrt{\pm i}\tan\sqrt{i\nu}=\frac{\sin\sqrt{2\nu}\mp\sinh\sqrt{2\nu}}{\sqrt{2}(\cosh\sqrt{2\nu}+\cos\sqrt{2\nu})},
\label{A5}
\end{equation}
we obtain
\begin{gather}
S=\frac{5a^{4}+\nu^{2}}{4a^{3}(a^{4}+\nu^{2})^{2}}+ \label{A6}\\
\frac{(a^{4}-2a^{2}\nu-\nu^{2})
\sinh\sqrt{2\nu}-(a^{4}+2a^{2}\nu-\nu^{2})\sin\sqrt{2\nu}}{%
2^{3/2}\nu^{3/2}(a^{4}+\nu^{2})^{2}(\cosh\sqrt{2\nu}+\cos\sqrt{2\nu})}.
\nonumber
\end{gather}
As a result, $P=F^{2}\ell ^{7}\omega ^{2}\eta S/\varepsilon ^{2}\lambda ^{2}$ reduces to Eq. (\ref{Q2}) in which $\chi=\nu/a^2$.  Equation (\ref{Q2}) can also be obtained  using Eq. (\ref{uloc}):
\begin{gather}
P=\frac{H^2_p\phi_0^2\omega}{2\lambda\varepsilon}\mbox{Im}\int_0^\ell\biggl[e^{-z/\lambda}+\frac{1}{q_\omega\lambda}\sinh(q_\omega z)- \nonumber \\
\frac{\cosh(q_\omega z)}{\cosh(q_\omega\ell)}\biggl(e^{-\ell/\lambda}+\frac{1}{q_\omega\lambda}\sinh(q_\omega\ell)\biggr)\biggr]\frac{e^{-z/\lambda}dz}{1-i\chi}.
\label{A7}
\end{gather}
Performing integration of Eq. (\ref{A7}) and neglecting terms $\sim e^{-\ell/\lambda}\ll 1$, we arrive at Eq. (\ref{Q1}) from which Eq. (\ref{Q2})
is obtained using Eq. (\ref{A5}).

For a film in a parallel field, $P$ can be calculated in the same way as above, but instead of $J(z) = (H_p/\lambda)e^{-z/\lambda}$
in a semi-infinite sample, we use $J(z)=(H_p/\lambda)\sinh(z/\lambda)/\cosh(d/2\lambda)$ where
$z=0$ is taken in the middle of the film. If $\ell=d/2$ (see Fig. 16), the solution for $u(z,\omega)$ which satisfies $u(0)=0$ and $u'(d/2)=0$
is then $u(z,\omega)=\sum_n A_n\sin(k_nz)$ where $k_n=\pi (2n+1)/d$, and $A_n$ is given by Eq. (\ref{A}). Here the formfactor $I_n$ which accounts for the spatial distribution of the rf driving force is replaced with $\tilde{I}_n$, where
\begin{gather}
\tilde{I}_n=\frac{4}{d}\int_0^{d/2}\frac{\sinh(z/\lambda)}{\cosh(d/2\lambda)}\sin(k_nz)dz \nonumber \\
\simeq \frac{4(-1)^nd}{\pi^2(2n+1)^2\lambda}, \qquad d\ll\lambda.
\label{A8}
\end{gather}
This expression was used to obtain Eq. (\ref{Pfilm}).

\section{Solution of thermal diffusion equation}
The partial Fourier transform $t_{\mathbf{p}}(z)=\int [T(\mathbf{r},z)-T_0]e^{i\mathbf{pr}}dxdy$ of Eqs. (\ref{tdif})-(\ref{bc2}) yields
\begin{eqnarray}
t_{\mathbf{p}}^{\prime \prime }-p^{2}t_{\mathbf{p}}=0,
\label{B1} \\
\kappa t_{\mathbf{p}}^{\prime }=-\tilde{P}_{\mathbf{p}},\qquad z=0
\label{B2} \\
\kappa t_{\mathbf{p}}^{\prime }=-t_{\mathbf{p}}\alpha_K,\qquad z=d
\label{B3}
\end{eqnarray}
where the prime denotes differentiation over $z$, and $\tilde{P}_{\mathbf{p}}=P_{\mathbf{p}}+(H_p^{2}/2)(\partial R_s/\partial T)t_{\mathbf{p}}(z)$.
Then Eq. (\ref{B2}) becomes:
\begin{equation}
\kappa t_{\mathbf{p}}^{\prime }+\Gamma t_{\mathbf{p}}=-P_{\mathbf{p}},\qquad z=0
\label{B4}
\end{equation}
where $\Gamma = (H_p^2/2)\partial R_s/\partial T$.
The solution of Eq. (\ref{B1}) is:
\begin{equation}
t_{\mathbf{p}}=A\cosh p(d-z)+B\sinh p(d-z),
\label{B5}
\end{equation}
where $p=|\mathbf{p}|$, and $A$ and $B$ are determined from the boundary conditions
(\ref{B3}) and (\ref{B4}):
\begin{gather}
Bp\kappa =A\alpha_K,
\label{B6} \\
p\kappa (A\sinh pd-B\cosh pd)-\nonumber \\
(A\cosh pd+B\sinh pd)\Gamma =P_{\mathbf{p}}.
\label{B7}
\end{gather}
Hence,
\begin{gather}
A=\frac{P_{\mathbf{p}}p\kappa }{p\kappa (\alpha_K-\Gamma)\cosh pd+(p^{2}\kappa ^{2}-\Gamma
\alpha_K)\sinh pd},
\label{B8} \\
B=\frac{P_{\mathbf{p}}\alpha_K}{p\kappa (\alpha_K-\Gamma)\cosh pd+(p^{2}\kappa ^{2}-\Gamma\alpha_K)\sinh pd}.
\label{B9}
\end{gather}
Substituting these formulas into Eq. (\ref{B5}), making the inverse Fourier transform and introducing the
dimensionless parameters $k=pd$, $\beta$ and $\gamma$ yields Eq. (\ref{tdis}).

Now we calculate $\delta T_s(\rho)$ at the outer surface $z=d$, for $r\gg d$.
In this case the main contribution to the integral in Eq. (\ref{tdis}) comes from
$k\ll 1$ for which the hyperbolic functions can be expanded in small $k$:
\begin{equation}
\delta T_s(\rho)=\frac{P_0\beta}{2\pi\alpha_K}\int_{0}^{\infty }\frac{J_0(k\rho)kdk}{b+ck^2}, \quad \rho\gg 1
\label{B10}
\end{equation}
where $b=\beta-\gamma(1+\beta)$ and $c=\beta^2+\beta(1-\gamma)/2-\gamma/6$. The integral in Eq. (\ref{B10}) equals
$K_0[\rho\sqrt{b/c}]/c$, Ref. \onlinecite{integr}, which reduces Eq. (\ref{B10}) to Eqs. (\ref{tin})-(\ref{alp}).
The value of $\delta T_s(0)$ is obtained from Eq. (\ref{tempo}) with $\rho=0$:
\begin{equation}
\delta T_s(0)=\frac{P_0\beta}{2\pi\alpha_K}\int_{0}^{\infty }\!\!\frac{k^{2}dk}{k\beta
(1-\gamma)\cosh k+(k^{2}\beta^{2}-\gamma)\sinh k}.
\label{B11}
\end{equation}
For $\beta\gg 1$, this integral can be done analytically by
introducing an auxiliary parameter $k_0$ such that $k_0\ll 1$ but $k_0^2\beta\gg 1$.
Then Eq. (\ref{B11}) splits into two parts:
\begin{gather}
\delta T_s(0)\cong \frac{P_0}{2\pi\alpha_K}\int_{0}^{k_{0}}\!\frac{kdk}{1-\gamma
+\beta k^{2}}+\frac{P_0}{2\pi\kappa d }\int_{k_{0}}^{\infty }\!\frac{dk}{\sinh
k}\notag \\
=\frac{q}{4\pi \kappa d}\ln \left[\frac{k_{0}^{2}\beta}{1-\gamma}\right]-\frac{q}{2\pi \kappa d}\ln \left[ \frac{k_0}{2}\right].
\label{B12}
\end{gather}
Here the parameter $k_0$ cancels out, resulting in
Eq. (\ref{dTs}).

Next we calculate $\langle\delta T^2\rangle$ in Eq. (\ref{dt2}):
\begin{equation}
\frac{\langle\delta T^2\rangle}{T_1^2}=\int_0^\infty\frac{(k\beta
 +\tanh k)^2e^{-k^2\varrho_0^2}kdk}{[k\beta (1-\gamma )+(k^2\beta^2-\gamma)\tanh k]^2},
\label{B13}
\end{equation}
where $T_1^2=H^4r_c^2\langle\delta R^2\rangle/8d^2\alpha_K^2$, and $\varrho_0=r_c/2d$. We calculate this
integral in the limit of $\beta\gg 1$ for which:
\begin{equation}
\frac{\langle\delta T^2\rangle}{T_1^2}=\int_0^\infty\frac{e^{-k^2\varrho_0^2}kdk}{[1-\gamma + k\beta\tanh k]^2}.
\label{B13}
\end{equation}
For the case of $\varrho_0\ll 1$ discussed in the text, the main contribution to the integral in Eq. (\ref{B13}) comes from $k\ll 1$
and $k\gtrsim 1$. Denoting the integral determined by the region of $k\ll 1$ as $I_1$, we have:
\begin{equation}
I_1=\int_0^\infty\frac{kdk}{[1-\gamma + \beta k^2]^2}=\frac{1}{2\beta(1-\gamma)},
\label{B14}
\end{equation}
where the upper limit was extended to $\infty$ to the accuracy of higher order terms $\sim \beta^{-2}\ll 1$. The part of the integral
in Eq. (\ref{B13}) determined by the region of $k>k_0\sim 1$ is
\begin{equation}
I_2=\frac{1}{\beta^2}\int_{k_0}^\infty\frac{dk}{k}e^{-k^2\varrho^2}\sim \frac{1}{\beta^2}\ln\frac{1}{k_0\varrho_0}.
\label{B15}
\end{equation}
For $\beta\gg 1$, the contribution of $I_1$ dominates, since $I_2/I_1\sim (1-\gamma)\ln(d/r_c)/\beta\ll 1$. Equations (\ref{B13}) and
(\ref{B14}) yield Eq. (\ref{dtb}).

%---------------------------------------------------------------bibliography

\end{document}